\documentclass[%
 reprint,
 amsmath,amssymb,
 aps,
]{revtex4-2}

\usepackage{graphicx}
\usepackage{dcolumn}
\usepackage{bm}
\usepackage{hyperref}
 \usepackage{xcolor}

\begin{document}

\preprint{APS/123-QED}

\title{\bf Trapped-atom Otto engine with light-induced dipole-dipole interactions}

\author{Chimdessa Gashu Feyisa\textsuperscript{1,2,4}}
\email{chimdessagashu@gmail.com}
\author{H. H. Jen\textsuperscript{1,2,3}}
\email{sappyjen@gmail.com}
\affiliation{\textsuperscript{1}Institute of Atomic and Molecular Sciences, Academia Sinica, Taipei 10617, Taiwan}
\affiliation{\textsuperscript{2}Molecular Science and Technology Program, Taiwan International Graduate Program, Academia Sinica, Taiwan}
\affiliation{\textsuperscript{3}Physics Division, National Center for Theoretical Sciences, Taipei 10617, Taiwan}
\affiliation{\textsuperscript{4}Department of Physics, National Central University, Taoyuan 320317, Taiwan}
\date{\today}

\begin{abstract}
Finite-time quantum heat engines are of practical relevance as they can generate finite-power, distinguishing them from ideal quasistatic engines typically used for theoretical purposes. However, these engines encounter energy losses due to quantum friction, which is particularly pronounced in many-body systems with non-trivial coherences in their density operator. Strategies such as shortcuts to adiabaticity and fast routes to thermalization have been developed although the associated cost requirements remain uncertain. In this study, we theoretically investigate the finite-time operation of a trapped-atom Otto engine with light-induced dipole-dipole interactions and projection measurements in one of the isochoric processes. The investigation reveals that appropriate control of dipole-dipole interactions of the working medium prompts engine operation upon interacting with the hot reservoir, while projection measurements and adjustment of the unitary driving protocols effectively reduce quantum friction to enhance finite-time engine performance compared to non-interacting and quasi-static counterparts. This setup presents a compelling avenue for further investigation of finite-time many-body quantum heat engines and provides an opportunity to explore the full potential of photon-mediated dipole-dipole interactions in their operations. 
\end{abstract}

\maketitle


\section{\label{I} Introduction}
Quantum thermodynamics is an exciting and rapidly growing field that lies at the intersection of various disciplines including quantum physics and thermodynamics. It seeks to unravel the fundamental principles governing energy transfer and conversion at the quantum level and to harnesses novel quantum phenomena to develop reliable quantum devices with efficient energy conversion and operational capabilities. One such frontier in this realm is the theoretical prediction and experimental realization of a quantum heat engine consisting spins \cite{ar1,ar2,ar3}, harmonic oscillators \cite{ar4,ar5}, trapped ions \cite{ar6,ar7,ar8,ar9}, and trapped atoms \cite{ar60,ar10,ar11,ar12} as its working medium. Traditionally, heat engine operation has relied on the presence of at least two heat reservoirs \cite{ar61,ar40}. However, in a surprising deviation, the quantum counterparts have demonstrated promising levels of efficiency even when operating with just a single non-thermal reservoir \cite{ar13,ar14,ar68}, or a thermal reservoir together with a projection measurement \cite{ar15,ar16} that can substitute the remaining heat bath \cite{ar8,ar9,ar16,ar17}. This remarkable phenomenon challenges conventional thermodynamic principles and opens up new possibilities for energy conversion in the quantum regime \cite{ar16,ar28,ar37,ar38}.

In a remarkable departure from conventional heat engines, a single-atom engine has been successfully realized in a linear Paul trap with tapered geometry \cite{ar6}. The precise control over the dynamics of trapped atoms also unlocks the exploration of their collective behavior arising from dipole-dipole interactions (DDIs), with implications spanning quantum information processing \cite{ar18,ar20}, quantum registers \cite{ar22,ar30}, and quantum heat engine design \cite{ar21,ar22}. These collective effects give rise to entangled states crucial for quantum communication and computation \cite{ar23,ar24}, as well as enabling the manipulation of atomic energy levels for implementing quantum gates and executing complex algorithms \cite{ar23,ar25}. Additionally, light-induced DDIs can impact the performance of quantum heat engines by modifying collective decays and frequency shifts \cite{ar24}, which can alter thermalization times and energy level spacings. This would potentially affect the efficiency of energy conversion processes and power generation in quantum heat engines. 

Despite the profound and intriguing physical phenomena under DDIs \cite{ar69}, their roles are relatively unexplored in engine operations, specifically considering interatomic distances and dipole polarization. A few studies have examined the collective advantages of (anti)ferromagnetic interactions \cite{ar25,ar26,ar27,ar28}, many-body effects at criticality \cite{ar21}, interacting Bose gases in a harmonic trap \cite{ar62}, and trapped ions with quasistatic conditions \cite{ar8}. Although the quasistatic mode of operation has practical limitations due to its vanishing output power, it provides a valuable theoretical framework for gaining a deeper understanding of the fundamental limits and capabilities of energy conversion processes in quantum systems \cite{ar43,ar63}. Besides, recent researches have examined finite-time operation of quantum heat engines with single-particle working media \cite{ar9,ar29,ar37,ar39}. While this advancement represents a shift towards more practical applications, natural systems are characterized by multiple interacting constituents whose collective effects play a vital role. In this regard, utilizing atoms with light-induced DDIs would be a more realistic approach that enables the tuning of interatomic distances and operation within finite-time frames. 

In this study, we explore a finite-time trapped-atom quantum Otto engine incorporating photon-mediated dipole-dipole interactions. We employ a quantum projection measurement technique \cite{ar15, ar16,ar17}, as well as the characteristics of open quantum systems \cite{ar24, ar30, ar31}, to handle the isochoric strokes in the engine. The projection measurement collapses the system to specific eigenstates, enabling controlled heat addition or removal and facilitating the generation of a heat gradient relative to the weakly coupled surrounding environment. We also explore finite-time engine performance with collective effects arising from the DDIs and many-body coherences that have been primarily limited to the single-particle case thus far \cite{ar9, ar29, ar37, ar39}. By identifying the optimal operating regime, particularly when the atoms are in close proximity, our results demonstrate simultaneous benefits of reduced thermalization time by more than sixty-fold compared to non-interacting working medium and quasistatic operation, as well as enhanced efficiency of work extraction, indicating improved performance. Besides, the performance of the engine is boosted through a collective frequency shift and projection measurement, both of which allow the engine to operate efficiently in the sudden unitary limit, provided that the working medium absorbs enough energy from the hot reservoir. This projection measurement, combined with appropriate unitary driving protocols, can mitigate coherence effects, which are the source of quantum friction. This further leads to a trade-off between power and efficiency in our system. Therefore, studying finite-time operations of trapped-atom engine with light-induced DDIs hold significant potential for advancing our understanding in quantum thermal machines and unlocking new possibilities in engine design with enhanced performance.

The rest of this paper is structured as follows. In Section \ref{II}, we introduce the model and Hamiltonian of the system. We next discuss the four thermodynamic processes that govern the trapped-atom Otto cycles in Section \ref{III}. In Section \ref{IV}, we examine finite-time performance of the engine under various scenarios such as finite-time thermalization, finite-time unitary driving, and a combination of both. Furthermore, we present a thorough analysis of the effect of light-induced DDIs on engine efficiency, output work and power. Finally, we provide concluding remarks in Section \ref{V}.
\section{\label{II} Hamiltonian of the system}
The working medium of the engine consists of two atoms coupled by light-induced DDIs \cite{ar69} and trapped in the Lamb-Dicke regime \cite{ar30,ar33,ar34}. In this regime, we truncate the motional degrees of freedom to the first-excited phonon mode, which is valid when the system is close to the motional ground state \cite{ar70,ar71,ar73}. This first-excited phonon state and the motional ground state enables precise control over the motion of the atoms and their interaction with external fields. Despite being a finite-level system, a phonon mode can serve as an effective heat sink for quantum systems \cite{ar8,ar9,ar30}. As a result, heat exchange takes place between the phonon mode and the atoms when they interact. Furthermore, both the working medium and the phonon mode weakly interact with the surrounding environment during the heating phase (see Section \ref{III} for the details). 

The working medium, phonon mode, and their interaction can be modeled by the total Hamiltonian given by
\begin{eqnarray}
\hat H(t)&=&\hat{H}_{\rm s}(t)+\hat{H}_{\rm ph}+\hat{H}_{\rm I},
\end{eqnarray}
where the Hamiltonian $\hat{H}_{\rm s}$ describes atomic system dynamics in the internal states, $\hat{H}_{\rm ph}$ represents phonon mode whose motion is confined to the Lamb-Dicke regime \cite{ar30,ar33,ar34}, and $\hat{H}_{\rm I}$ characterizes atom-phonon couplings. The Hamiltonians can be written as ($\hbar=1$) \cite{ar8,ar24,ar30}
\begin{eqnarray}
\hat H_{\rm s}(t)&=&g\sum^2_{i=1}\hat\sigma^{x}_{i}+B(t)\sum^2_{i=1}\hat\sigma^{z}_{i}\nonumber\\&+&\sum^2_{i\neq j}\sum^2_{i,j=1}\Omega_{ij}\hat\sigma^\dag_{i}\hat\sigma_{j}\label{1},\\
\hat{H}_{\rm ph}&=&\omega\hat a^\dag\hat a,\\
\hat H_{\rm I}&=&\sum^2_{i=1}\chi_i(\hat a\hat\sigma^\dag_i+\hat a^\dag\hat\sigma_i).
\end{eqnarray}

In these equations, the driving fields $B(t)$ and $g$ act along the longitudinal and transverse directions, respectively, while the interaction strength $\chi_i$ governs the coupling between atoms and phonon mode. The phonon mode is described by annihilation $\hat{a}$ and creation $\hat{a}^\dagger$ operators with a harmonic trap frequency $\omega$, while each atom is characterized by the Pauli matrices satisfying the relationships $\hat\sigma_i^z = \hat\sigma_i^\dag\hat\sigma_i - \hat\sigma_i\hat\sigma_i^\dag$, $\hat\sigma_i^x = \hat\sigma_i^\dag + \hat\sigma_i$ and $\hat\sigma_i^\dag=\big(\hat\sigma_i\big)^\dag$, where $\hat\sigma_i$ represents the transition from the excited state $|e_i\rangle$ to the ground state $|g_i\rangle$. Additionally, $\Omega_{ij}$ is the frequency shift due to the photon-mediated DDIs and given by \cite{ar24,ar30}
\begin{eqnarray}
\Omega_{12}&=&\frac{3\Gamma}{4}\bigg[-\big(1-{\rm cos^2}\theta\big)\frac{{\rm cos}(\xi)}{\xi}\nonumber\\&+&(1-3{\rm cos^2}\theta\big)\big(\frac{{\rm sin}(\xi)}{\xi^2}+\frac{{\rm cos}(\xi)}{\xi^3}\big)\bigg],\label{cf}
\end{eqnarray}
where $\Gamma$ represents the effective decay rate of excited atoms, and $\theta={\rm cos^{-1}}(\hat \mu.\hat r_{ij})$ is the dipole polarization angle with $\hat{\mu}$ and $\hat{r}_{ij}$ being unit vectors in the directions of the dipole moment and inter-atomic spacing, respectively. The relative distance between atoms is quantified by a dimensionless parameter $\xi=|{\vec k}||{\vec r}_{ij}|$, in which the wave-vector $|{\vec k}|$ is calculated as $2\pi/\lambda_0$ with a wavelength of light $\lambda_0$.

Trapped atoms in the Lamb-Dicke regime can be attained under the conditions of $\omega \gg \Gamma$ and $\eta_d \ll 1$, where $\eta_d$ represents the Lamb-Dicke parameter defined as $\eta_d = k_{\text{eff}}/\sqrt{2m\omega}$, with $m$ being the atomic mass and $k_{\text{eff}}$ denoting the effective wave vector tailored to the specific trapping technique under consideration \cite{ar30}. The physical system being considered offers great flexibility, enabling independent control of the working medium through the fields and adjustment of inter-atomic distances. For instance, when the parameter $g$ vanishes, the energy levels of the system Hamiltonian do not intersect during finite-time unitary processes. In such cases, the distinction between finite-time and quasistatic operations of the engine becomes irrelevant \cite{ar1}. This is because the power output remains zero, while the engine efficiency remains the same as the quasistatic value. However, when $g\neq0$, the system Hamiltonians at different times, $t_a$ and $t_b$, do not commute with each other, $\big[\hat H_s(t_a), \hat H_s(t_b)\big]\neq 0$, leading to the emergence of quantum friction during finite-time operation of the engine \cite{ar1,ar2,ar36}. 

Furthermore, the measurement protocol presents an intriguing and easily implementable feature worthy of investigation \cite{ar15,ar17}. It effectively mitigates the detrimental effects of quantum friction and ensures the smooth operation of the engine, even in the face of sudden unitary dynamics \cite{ar37,ar38}. Consequently, it paves the way for exciting new possibilities in the experimental realization of quantum heat engines. Our study also investigates the influences of two-atom coherence on engine performance during finite thermalization and unitary driving protocols, expanding upon previous research mainly focused on the single atom working medium \cite{ar9,ar29,ar37,ar39}. In the next section, we will delve more into the engine cycle of trapped atoms with photon-mediated DDIs.
\begin{figure}
\begin{center}
\resizebox{0.45\textwidth}{!}{%
\includegraphics{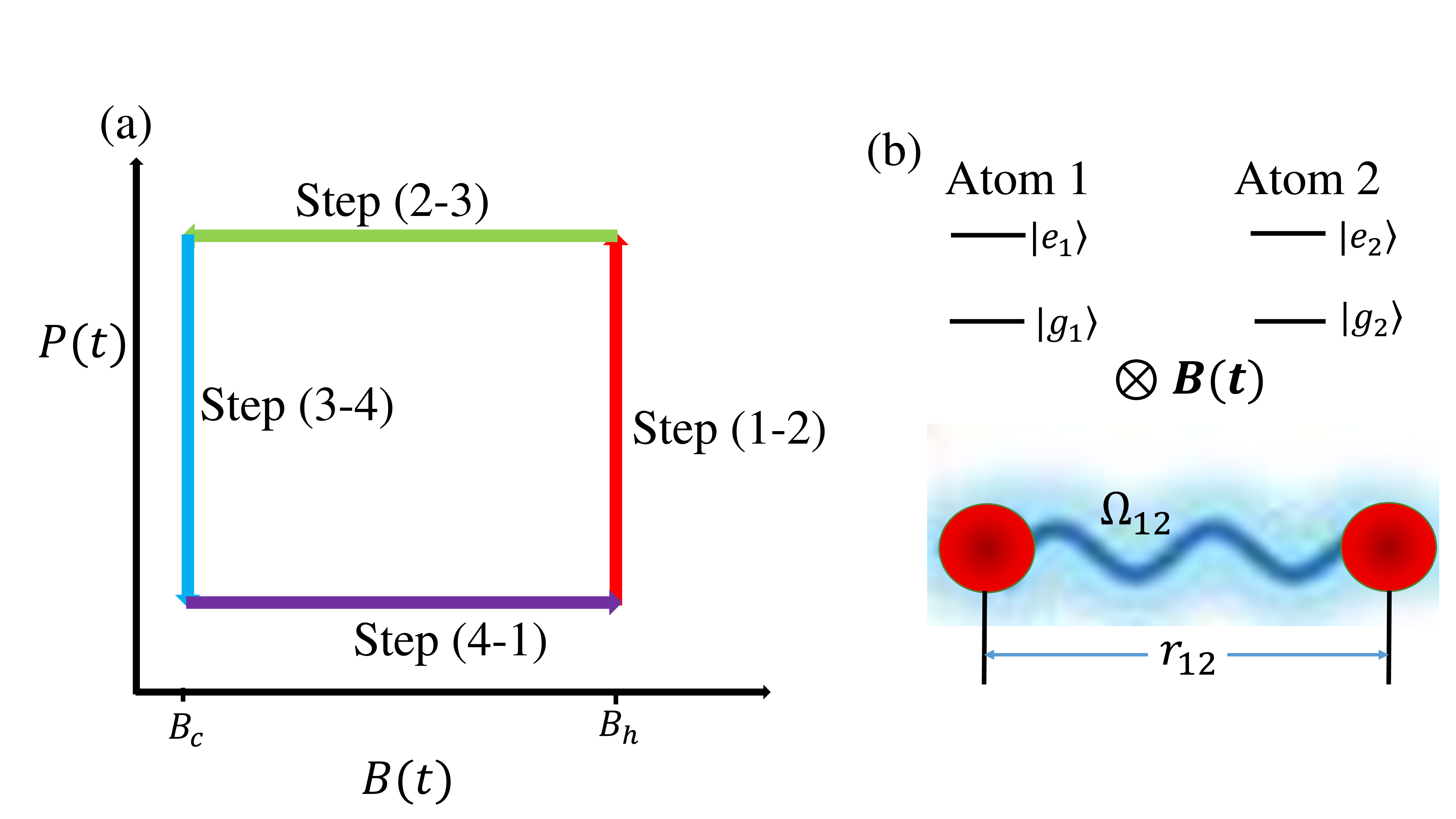}}
\caption {Probability distributions versus control parameter of the trapped-atom quantum Otto engine processes. (a) In step (1-2), the working medium which is initially in state $\hat{\rho}(0)$ interacts with a hot reservoir for a finite-time $t \in [0, t_1]$, while a magnetic field is fixed at $B_h$. The state $\hat{\rho}_1(t)$ of the system is determined by Eq. (\ref{mom}) at each time during the heating process. In step (2-3), the working medium undergoes a unitary expansion over a finite-time interval $t \in [t_1, t_2].$ The system's state changes to $\hat{\rho}_2(t)$, and work $W_{23}$ is done by changing the external magnetic field from $B_h$ to $B_c$. In step (3-4), the working medium rejects heat to the phonon mode by projecting it onto the initial state, while maintaining a magnetic field at $B_c$. This heat rejection process is made instantaneous as $t_2 \approx t_3$. In the last step (4-1), the working medium is compressed in a finite-time unitary process for $t \in [t_3, t_4]$. The system's state changes to $\hat{\rho}_4(t)$, and work $W_{41}$ is performed. (b) The system consists of two atoms, labeled as Atom 1 and Atom 2, subjected to time-dependent magnetic fields denoted by $B(t)$. The atoms are coupled through photon-induced dipole-dipole interactions $\Omega_{12}$ and coupling field determined by constant $g$ in Eq. (\ref{1}).}\label{f1}
\end{center}
\end{figure}
\section{\label{III} Quantum Otto engine cycle with trapped atoms}
We consider a four-stroke Otto cycle with two unitary and two isochoric processes \cite{ar48,ar49}. This engine cycle allows for the transfer of heat and work done separately during each thermodynamic stroke \cite{ar48}. In our system, one of the isochoric steps is replaced by an instantaneous projection measurement \cite{ar9,ar15,ar16}. The hot reservoir is an external environment that is weakly coupled to the working medium, which can be characterized by the theory of open quantum systems \cite{ar24} to handle this interaction and account for finite thermalization time and collective effects from light-induced DDIs \cite{ar69}. Starting from an arbitrary initial state of the two atoms, we perform an isochoric heating step (1-2) to initiate the engine cycle in Fig. \ref{f1}(a). If the atoms are initially in the ground state, they absorb heat energy from the hot reservoir and the temperature of the working medium increases. In contrast, initially excited atoms are cooled by releasing energy to the reservoir. The choice of the initial state determines whether energy is absorbed or rejected as heat, influencing the temperature of the working medium \cite{ar8}.

Our engine operation focuses on exciting ground state atoms during step (1-2) via heat absorption from the hot reservoir, as depicted in Fig. \ref{f1}(a). This is followed by a unitary step (2-3), which enables work performance if sufficient heat has been absorbed from the hot reservoir \cite{ar50,ar51}. The extraction of work is facilitated by subjecting the interacting atoms to a time-dependent magnetic field, as illustrated in Fig. \ref{f1}(b). The subsequent engine cycle involves a projective measurement acting as a cooling step (3-4). This step rejects a portion of heat to the phonon mode, which serves as a cold bath for the two atoms in our setup \cite{ar16,ar52}. Following the cooling process, we evolve the system to complete the engine cycle by reversing the protocol used in step (2-3). It is worth noting that the probability of finding atoms in the ground state after this compression step (4-1) can be nearly unity to ensure that the engine cycle is complete with a minimal energy dissipation due to finite-time operation. More details about atomic excitations in the unitary strokes have been explained in the next section and analyzed in Section \ref{IV} B and C as well as in Appendix \ref{B}. 
\subsection{Isochoric heating process}
The working substance of a quantum heat machine can be coupled thermally to one or more heat reservoirs consisting of an infinite number of harmonic oscillators \cite{ar4, ar24, ar40}. After a sufficiently long thermalization time, ideally infinite, this thermal coupling brings the system into equilibrium with a bosonic hot reservoir characterized by an inverse temperature $\beta=1/(k_{B}T)$ with Boltzmann constant $k_B$ and an average number of photons $\bar{n}_{th}=(\exp({2\beta B_h})-1)^{-1}$ with a magneic field $B_h$ \cite{ar24,ar40,ar72}. Under these conditions, the system can be effectively described by a Gibbs state \cite{ar42, ar58}. Conversely, for a short thermalization time, the working medium remains out of thermal equilibrium, and quantum coherence will be retained in the final density matrix elements \cite{ar29}. To analyze this finite-time thermalization process, we employ the framework of open quantum theory, wherein the dynamics of the working medium is governed by a Lindblad master equation written as \cite{ar24, ar30, ar31, ar32}
\begin{eqnarray}
 \frac{d\hat\rho(t)}{dt}&=&-i[\hat H(0),\hat\rho(t)]+\sum_{s=\pm}\frac{\gamma_{s}}{2}(\bar {n}_{th}+1){\cal{L}_{\substack{s}}}\nonumber\\&+&\sum_{s=\pm}\frac{\gamma_{s}\bar{n}_{th}}{2}{\cal{L}^{\dag}_{\substack{s}}}\label{mom}.
\end{eqnarray}

In this equation, the magnetic field is set to a fixed value $B_h$ to maintain a time-independent system Hamiltonian, while the Lindblad operator is given by ${\cal{L}_{\substack{s}}}[\hat\rho(t)]=2\hat\sigma_{s}\hat\rho(t)\hat\sigma^\dag_{s}-\hat\sigma^\dag_{s}\hat\sigma_{s}\hat\rho(t)-\hat\rho(t)\hat\sigma^\dag_{s}\hat\sigma_{s}$, in which $\hat\sigma^\dag_{\pm}=\frac{\hat\sigma^\dag_{1}\pm\hat\sigma^\dag_{2}}{\sqrt{2}}$, $\hat\sigma_{\pm}=\frac{\hat\sigma_{1}\pm\hat\sigma_{2}}{\sqrt{2}},$ and $\gamma_{\pm}=\gamma\pm\gamma_{12}$. Here $\gamma_{12}$ ($\gamma=\gamma_{11}=\gamma_{22}$) represents the collective (individual) decay of the two atoms due to direct interaction with the surrounding environment, and the collective decay is written in the form \cite{ar24,ar30}
\begin{eqnarray}
\gamma_{12}&=&\frac{3\Gamma}{2}\bigg[\big(1-{\rm cos^2\theta}\big)\frac{{\rm sin}(\xi)}{\xi}\nonumber\\&+&(1-3{\rm cos^2\theta}\big)\big(\frac{{\rm cos}(\xi)}{\xi^2}-\frac{{\rm sin}(\xi)}{\xi^3}\big)\bigg].
\end{eqnarray}
This collective decay $\gamma_{12}$ as well as frequency shift $\Omega_{12}$ given in Eq. (\ref{cf}) emerge from system-reservoir interactions, and they satisfy Kramers-Kronig relation and sustain causal relations required in electrodynamics theory. 

In the isochoric processes, the internal energy of the working medium can solely be altered by the exchange of heat with a reservoir. The amount of heat transfer to the working medium during the heating process is quantified as \cite{ar1,ar63,ar29,ar43} $Q_h=U_1-U_0,$ where $U_0={\rm Tr_{ph}}\big[\hat\rho(0)\hat{H}_s(0)\big]$ and $U_1={\rm Tr_{ph}}\big[\hat\rho_1(t)\hat {H}_s(0)\big]$ represent the energy of the working medium before and after the isochoric heating process, respectively. Here, $\hat{\rho}(0)$ denotes the initial state and $\hat{\rho}_1(t)$ denotes the final state of the working medium as described in Eq. (\ref{mom}) for $t \in [0, t_1]$. If this thermalization time is longer than the relaxation time of the system, the system is fully thermalized to the hot bath \cite{ar29}. Conversely when the relaxation time is on the order of or greater than the thermalization time, the state $\hat\rho_1(t)$ can carry some amount of quantum feature which will be transferred to the next engine cycle. The deviation of the working medium from equilibrium condition during the heating process can be analyzed using fidelity of the first stroke calculated as \cite{ar1,ar56} $${\cal F}_{12}=\text{Tr}\sqrt{\sqrt{\hat{\rho}^{t_1\rightarrow\infty}_1}\hat{\rho}_1(t)\sqrt{\hat{\rho}^{t_1\rightarrow\infty}_1}},$$ where $\hat{\rho}^{t_1\rightarrow\infty}_1 = \exp(-\beta\hat{H}_s(B_h))/Z_h,$ with a partition function $Z_h=\text{Tr}\big[\exp(-\beta\hat{H}_s(B_h))\big],$ represents a thermal state reached at the end of the isochoric heating step (1-2)
\subsection{Unitary expansion process}
In unitary processes, the Hamiltonian of the working medium is externally modified akin to pistons of classical heat engines \cite{ar1}. These processes can occur suddenly $\tau\rightarrow0$, in finite-time $\tau$ or quasi-statically $\tau\rightarrow\infty$ \cite{ar5,ar38}. In this work, we encompass both sudden and quasi-static unitary processes as the extreme cases. A notable characteristic of finite-time unitary processes is the generation of quantum internal friction, stemming from the non-commutativity of the Hamiltonians of the working medium at different times. This friction arises due to the production of finite quantum coherence during short driving intervals and the irreversibility of the system caused by entropy production \cite{ar5,ar44,ar45}. The coherence between non-degenerate energy eigenstates can have a detrimental effect on engine efficiency \cite{ar37,ar46}. However, in specific scenarios, it can also serve as a quantum lubricant, mitigating quantum friction when considering both finite-time thermalization and unitary driving \cite{ar29}. This occurs due to the interference between quantum coherence generated during the finite-time unitary driving process and the remaining coherence sustained in the finite-time thermalization step in the engine cycle. Conversely, coherence generated between degenerate energy levels boosts engine performance \cite{ar46}. 

In our setup, we can implement the unitary expansion stage by modifying various parameters such as the time-dependent driving field $g$, the magnetic field $B$, or the coupling parameter $\Omega_{12}$, either individually or in combination. Here, we consider a time-dependent magnetic field \cite{ar9,ar40} $B(t)=B_h+(B_c-B_h)t/\tau$ that changes from $B_h$ at $t=0$ to $B_c$ at $t=\tau$ for the expansion step, which occurs on a shorter time scale compared to the dissipation rates. In this scenario, we can ignore heat exchange between the system and the environment, enabling us to evolve the working medium through unitary dynamics written as
\begin{eqnarray}
 \frac{d\hat\rho(t)}{dt}&=&-i[\hat H(t),\hat\rho(t)] \label{me2},
\end{eqnarray}
and work done on the engine by the driving agent can be determined as \cite{ar1,ar63,ar29,ar43} $W_{23}=U_{2}-U_{1},$ where $U_2={\rm Tr_{ph}}\big[\hat{\rho}_2(t)\hat{H}_s(\tau)\big]$ is the internal energy of the system during the evolution. Here, $\hat{\rho}_2(t)$ is calculated from Eq. (\ref{me2}) for the time interval $t\in[t_1,t_2]$, and $\tau = t_2 - t_1$ represents the duration of the expansion process. 

Moreover, the energy dissipated by quantum friction in this process can be explicitly determined by \cite{ar47} $W_{23}^{\text{fri}} = W^{\text{nad}}_{23} - \Delta F_{\text{exp}}$, where $W^{\text{nad}}_{23}$ is the adiabatic work done in the expansion process, and $\Delta F_{\text{exp}}$ is the associated change in free energy. The partition functions are used to calculate the free energies before and after the unitary expansion. The dissipated work can also be estimated using relative entropy \cite{ar9,ar47} and fidelity \cite{ar1}, as these quantities measure the distance between non-equilibrium and equilibrium states. Relative entropy focuses on quantifying the difference from equilibrium, while fidelity provides a measure of how well a non-equilibrium state aligns with an equilibrium state. In this work, we employ fidelity to assess the proximity of a non-equilibrium state to an equilibrium state. For the expansion stage, fidelity can be written in the form \cite{ar1,ar56} $${\cal F}_{23}=\text{Tr}\sqrt{\sqrt{\hat{\rho}^{\tau\rightarrow\infty}_2}\hat{\rho}_2(t)\sqrt{\hat{\rho}^{\tau\rightarrow\infty}_2}},$$ where $\hat{\rho}^{\tau\rightarrow\infty}_2 = \exp(-\beta\hat{H}_s(\tau))/Z$ with a partition function $Z=\text{Tr}\big[\exp(-\beta\hat{H}_s(\tau))\big]$ is a thermal state obtained at the end of the unitary expansion. We recall that the dissipated work originates from the non-commutativity of the system Hamiltonians at different times \cite{ar5,ar29} and can be ignored when either of the driving fields are off or when the adiabatic condition is satisfied \cite{ar1}.
\subsection{Isochoric cooling process}
In this thermodynamic process, we harness the concept of projection measurement to precisely manipulate the state of the working medium \cite{ar15,ar17}. This technique relies on a fundamental principle of quantum mechanics that provides information about the quantum state of a system through measurement. Notably, this projection measurement is instantaneous and devoid of any delay \cite{ar15,ar16,ar37,ar38}. The choice of projected states plays a critical role in determining the operational mode of the machine, whether it functions as an engine or a refrigerator \cite{ar8}. The heat released from the working medium in this step can be quantified using the expression \cite{ar8,ar29,ar43} $Q_c=U_3-U_2$, where the energy of the working medium modifies to $U_3={\rm Tr_{ph}}\big[\hat{\rho}_3(t)\hat{H}_s(\tau)\big].$ Here, $\hat{\rho}_3(t)$ is determined based on the chosen measurement basis while the duration of cooling process is made instant as $t_c=t_3-t_2=0$.   
\subsection{Unitary compression process}
In the unitary compression stage we perform a time reversed protocol of the unitary expansion, specifically we compress the working medium by changing the magnetic field back to $B_h$ to resume the next engine cycle. Therefore, the state of the working medium $\hat{\rho}_4(t)$ and the work done $W_{41} = U_{4} - U_{0}$, with $U_{4} = {\rm Tr_{ph}}\big[\hat{\rho}_4(t)\hat{H}_s(\tau)\big]$, are determined using Eq. (\ref{me2}) under time-dependent $\hat{H}(t)$ within the interval $t\in[t_3,t_4]$. We assume the time elapsed for both unitary dynamics is equally synchronized to $\tau$, such that $t_4 - t_3 = \tau$. Moreover, the finite compression step dissipates energy \cite{ar47} $W^{\text{fri}}_{41}=W^{\text{nad}}_{41} - \Delta F_{\text{comp}}$, where $W^{\text{nad}}_{41}$ is the adiabatic work done and $\Delta F_{\text{comp}}$ is the free energy change in the process. We analyze the deviation of the final state from the initial ground state, from which the engine starts, using fidelity \cite{ar1,ar56} written in the form $${\cal F}_{41}=\text{Tr}\sqrt{\sqrt{\hat{\rho}(0)}\hat{\rho}_4(t)\sqrt{\hat{\rho}(0)}}.$$ All simulations are carried out using QuTip packages \cite{ar57}.
\section{\label{IV} Performance of trapped-atom engine}
\subsection{Finite thermalization and adiabatic case}
\begin{figure}
\begin{center}
\hspace{-2.75cm}{(a)\hspace{3.75cm}(b)}\\
\includegraphics[width=.45\textwidth]{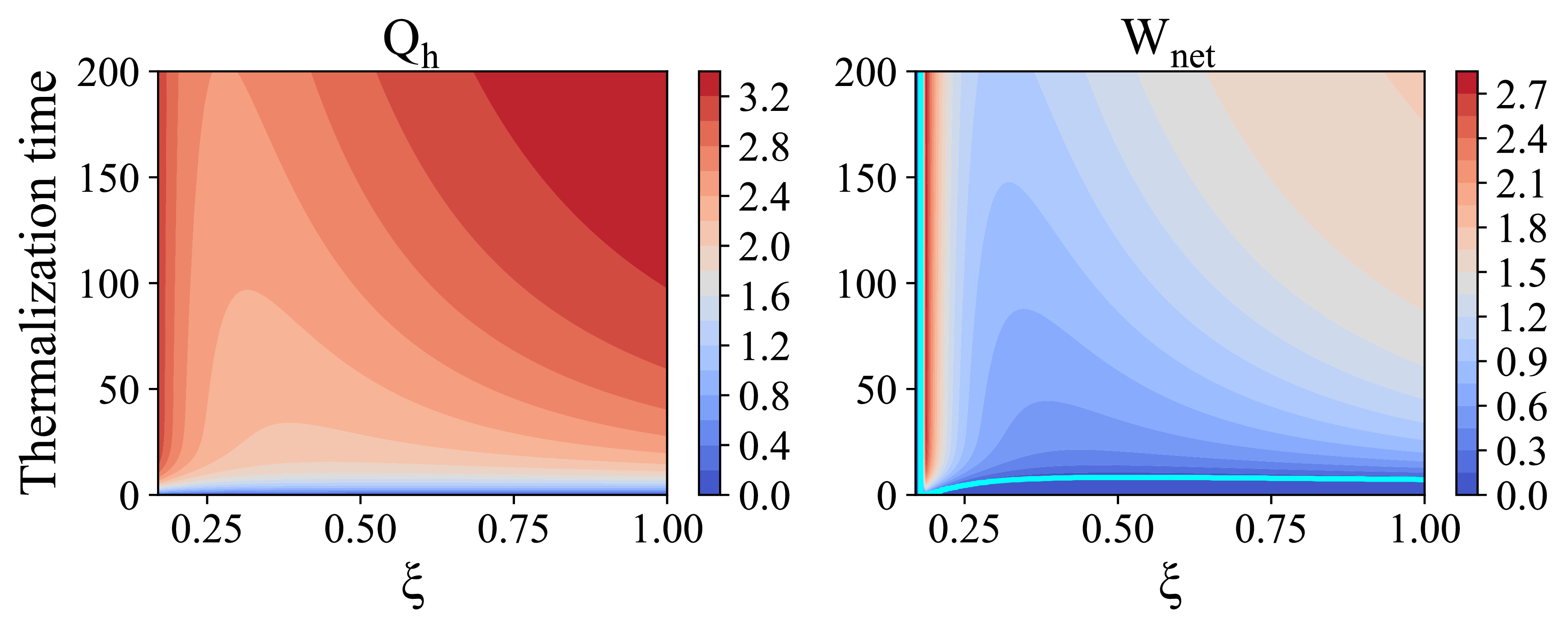}\\
\hspace{-2.75cm}{(c)\hspace{4cm}(d)}\\
\includegraphics[width=.45\textwidth]{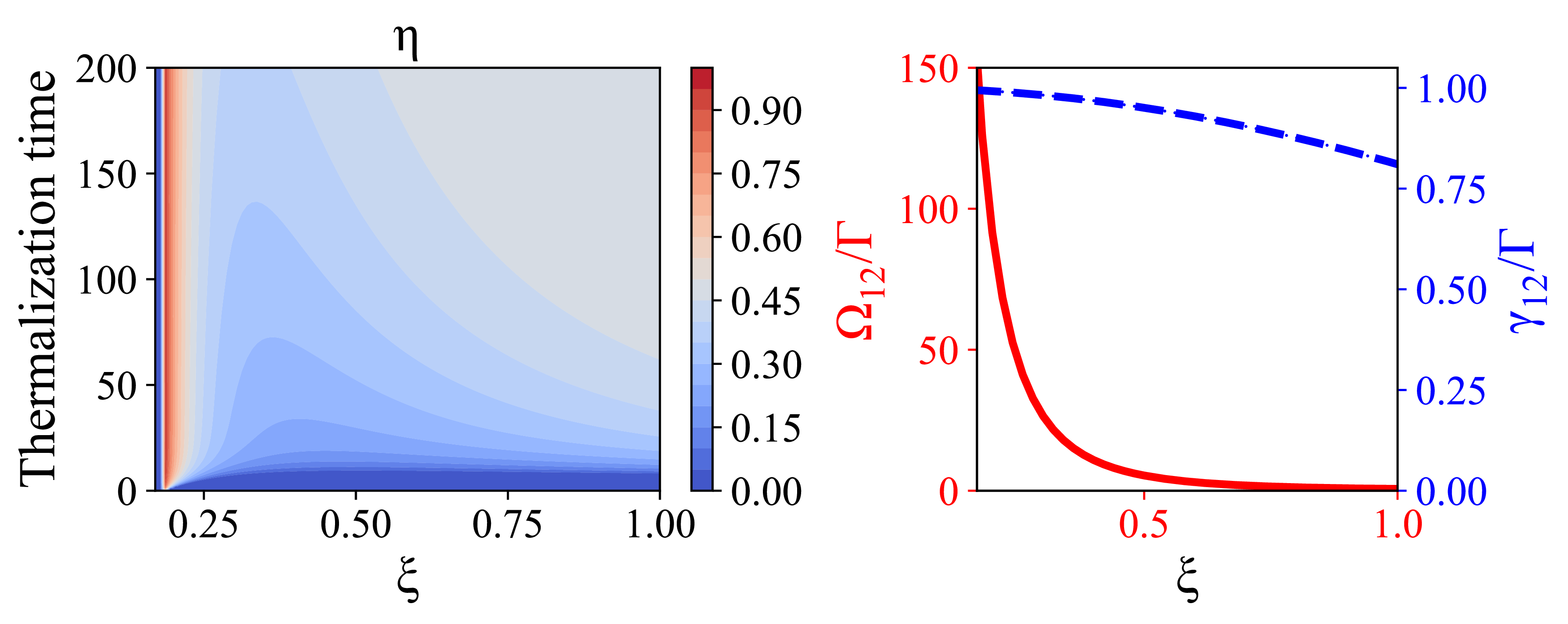}\\
\hspace{-2.75cm}{(e)\hspace{3.75cm}(f)}\\
\includegraphics[width=.45\textwidth]{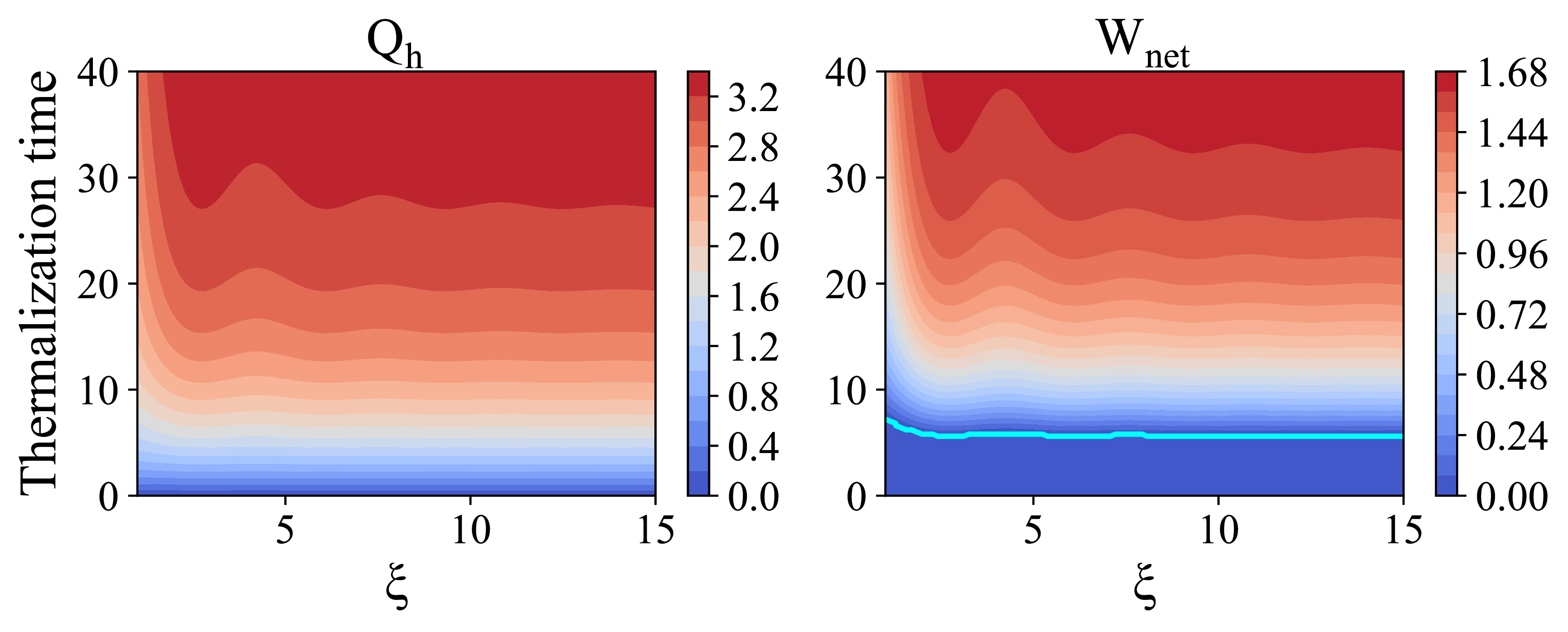}\\
\hspace{-2.75cm}{(g)\hspace{4cm}(h)}\\
\includegraphics[width=.45\textwidth]{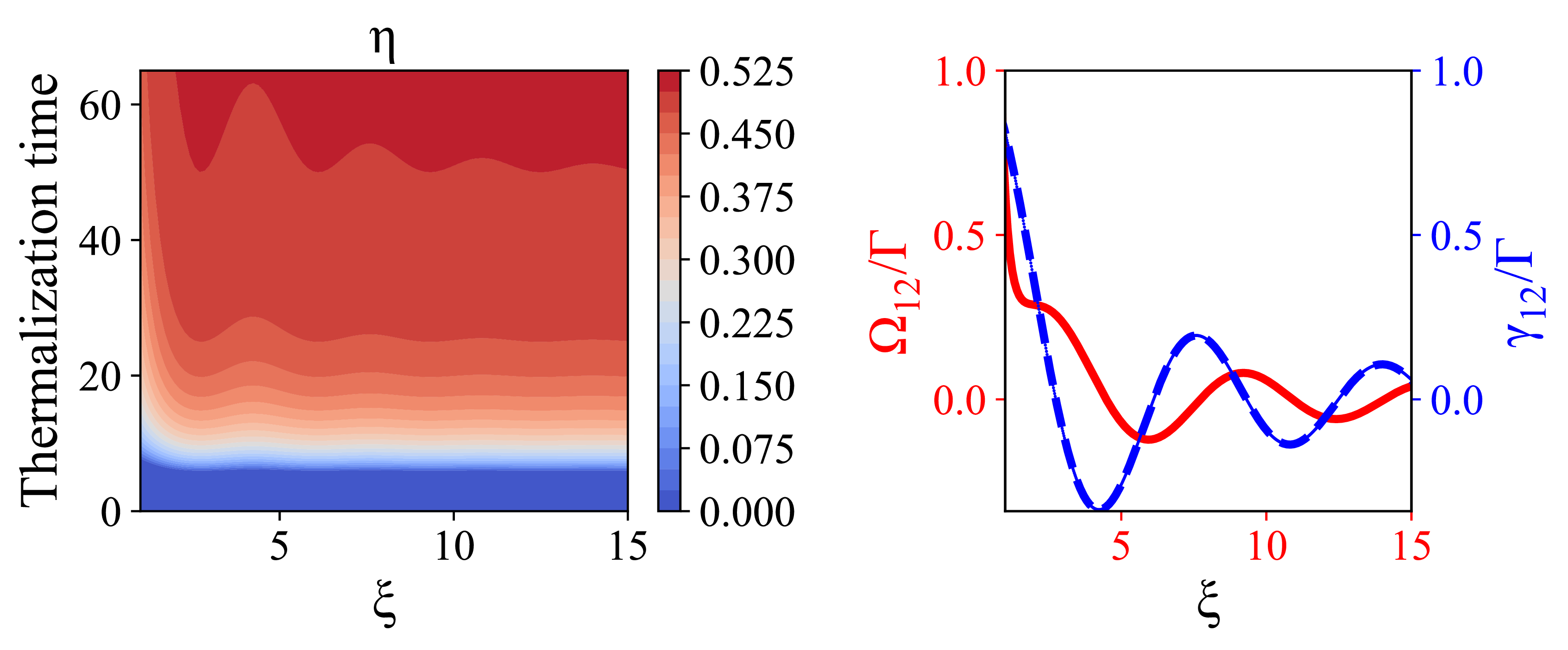}
 \caption{Trapped-atom engine operation for different interatomic distances ($\xi\in[0.175,15]$) and thermalization times. Panels (a) to (d) display heat absorbed from hot reservoir, net work done by the engine (thick cyan line marks vanishing net work output), engine efficiency, and light-induced DDIs, respectively, for $\xi\in[0.175,1]$. Panels (e) to (h) show the corresponding results for $\xi\in[1,15]$. The remaining parameters are given by $\theta=\pi/2$, $g=0.2\omega$, $B_{c(h)}=5(10)\omega$, $\chi_{1}=\chi_2=0.04\omega$, $\Gamma=0.1\omega$, $\bar{n}_{th}=0.1$, and $\omega=1$.}\label{f2}
\end{center}
\end{figure}

We consider a finite-time thermalization process of the working medium during the isochoric heating step, while ensuring that the unitary processes remain adiabatic throughout the engine operation \cite{ar9,ar66,ar67}. Adiabatic conditions pertain to a quasi-static process wherein the magnetic field slowly transitions from $B_h$ to $B_c$ and vice versa during the second and fourth strokes \cite{ar8,ar48}. Throughout these strokes, the states of the working medium remain unchanged to ensure that there is no energy dissipation due to entropy change \cite{ar5,ar36}. This approach allows us to carefully manage the thermalization process of the working medium, optimizing its performance. 

The performance of the trapped-atom engine is evaluated in terms of efficiency given by $\eta=W_{net}/Q_h$, where $W_{net}=W_{23}+W_{41}$ and $Q_h$ represent the net work done and heat absorbed by the engine, respectively (see Section \ref{III} for the details). In order to achieve operational effectiveness as a heat engine and produce a net work output of $W_{net}$, the thermal device must satisfy the following criteria \cite{ar63}: heat absorption from a hot reservoir ($Q_h>0$), heat rejection to a cold reservoir ($Q_c<0$), and a condition where the magnitude of absorbed heat exceeds the magnitude of rejected heat ($|Q_h| > |Q_c|$). This regime is achieved when interparticle distance $\xi>0.176$ ($r_{12}>0.028\lambda_0$), and when sufficient time is allocated for the heating process to ensure that the working medium receives enough heat to generate work output effectively. Full thermalization and adiabatic conditions correspond to the quasi-static and ideal operation of the engine, as discussed in Ref. \cite{ar8} for the case of a trapped-ion engine with spin-exchange interaction. In this case, the spin-exchange interaction enhances the engine efficiency beyond the non-interacting efficiency limit $\eta=1-B_c/B_h$ for $g=0$ and until the interaction strength is comparable to the magnetic field $B_h$. However, spin-exchange interaction stronger than $B_h$ would abruptly distort the atomic energy levels, rendering the system ineffective as an engine. 

We first examine the effect of different atomic configurations on the engine performance by choosing perpendicular dipole orientation and range of interatomic distances $r_{12}\in[0.028\lambda_0,2.39\lambda_0]$ as depicted in Figs. \ref{f2}(a-h). Figs. \ref{f2}(a-d) present the results for short-distance $\xi$, and we continue the analysis for a longer distance $\xi$ in Figs. \ref{f2}(e-h). Specifically, the former range covers interatomic distances up to approximately $0.15\lambda_0$, where light-induced DDIs are stronger compared to the latter case, which encompasses interatomic distances up to $2.4\lambda_0$. In these specific parameter regimes, the behavior of the trapped-atom engine varies depending on the strength of the light-induced DDIs, as can be seen in Figs. \ref{f2}(a-c) and \ref{f2}(e-g). For short interatomic separations, $\xi<0.176$, the system does not operate as a heat engine since the working medium fails to extract work output from the hot reservoir. In this regime the thermal device absorbs heat from both hot and cold reservoirs, thus, a shorter interatomic separation is not available for industrial application under the chosen parameter spaces and protocols. 

The thermalization time required to initiate engine operation depends significantly on the interatomic distance, as depicted in Figs. \ref{f2}(a-c) and \ref{f2}(e-g). When the atoms are in close proximity, the dominant factor is the collective frequency shift between them, enabling the engine to achieve near-perfect efficiency and immediate operation upon interaction with the hot reservoir. Specifically, the thermalization time required for the working medium to generate workout is approximately $0.1/\omega$ for $\xi=0.19$, whereas it is nearly $6/\omega$ for $\xi=15$ (see Figs. \ref{f2}(b) and \ref{f2}(f)). In this scenario, the working medium with a relatively short interatomic distance efficiently converts a substantial amount of heat absorbed from the hot reservoir into useful work output within a brief time, which also has a positive impact on the power output. 

However, for interatomic separation of $\xi\in[0.176,1]$, the frequency shift decreases rapidly while the collective dissipation changes marginally, leading to a prolonged thermalization time that affects the engine operation. This prolonged thermalization can be attributed to collective dissipation, impeding heat transport between the working medium and the reservoirs. As a result, energy dissipation increases, resulting in a drop in efficiency and work output, as indicated in Figs. \ref{f2}(a), \ref{f2}(b), and \ref{f2}(c). When a system avoids complete thermalization, it implies that it retains quantum information in its state, which in turn affects its dynamics, and enables the emergence of unique and stable quantum phases \cite{ar58}. It is worth noting that the frequency shift remains small and finite as the interatomic distance increases (see Figs. \ref{f2}(d) and \ref{f2}(h)). Further details of the thermalization process including dipole orientation, and short-time and long-time behaviors of thermodynamics quantities are discussed in Appendix \ref{A}.
\begin{figure}
\begin{center}
\hspace{-2.75cm}{(a)\hspace{3.75cm}(b)}\\
\includegraphics[width=.5\textwidth]{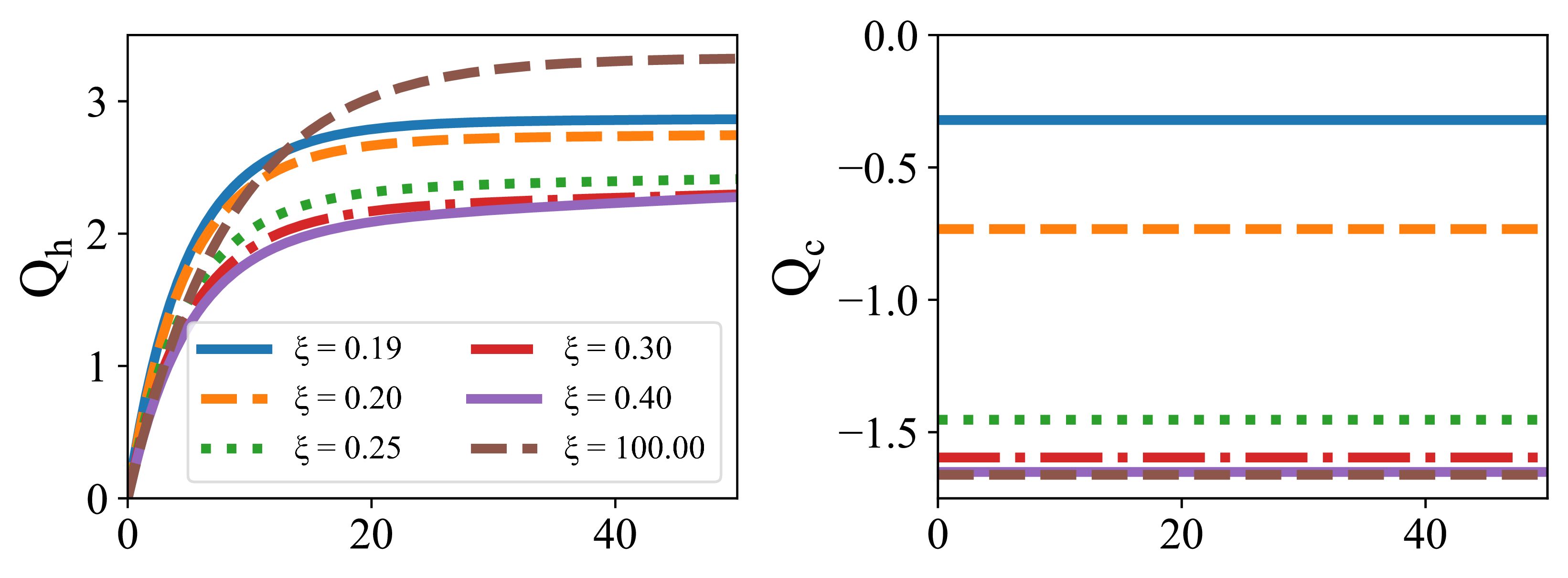}\\
\hspace{-2.75cm}{(c)\hspace{4cm}(d)}\\
\includegraphics[width=.5\textwidth]{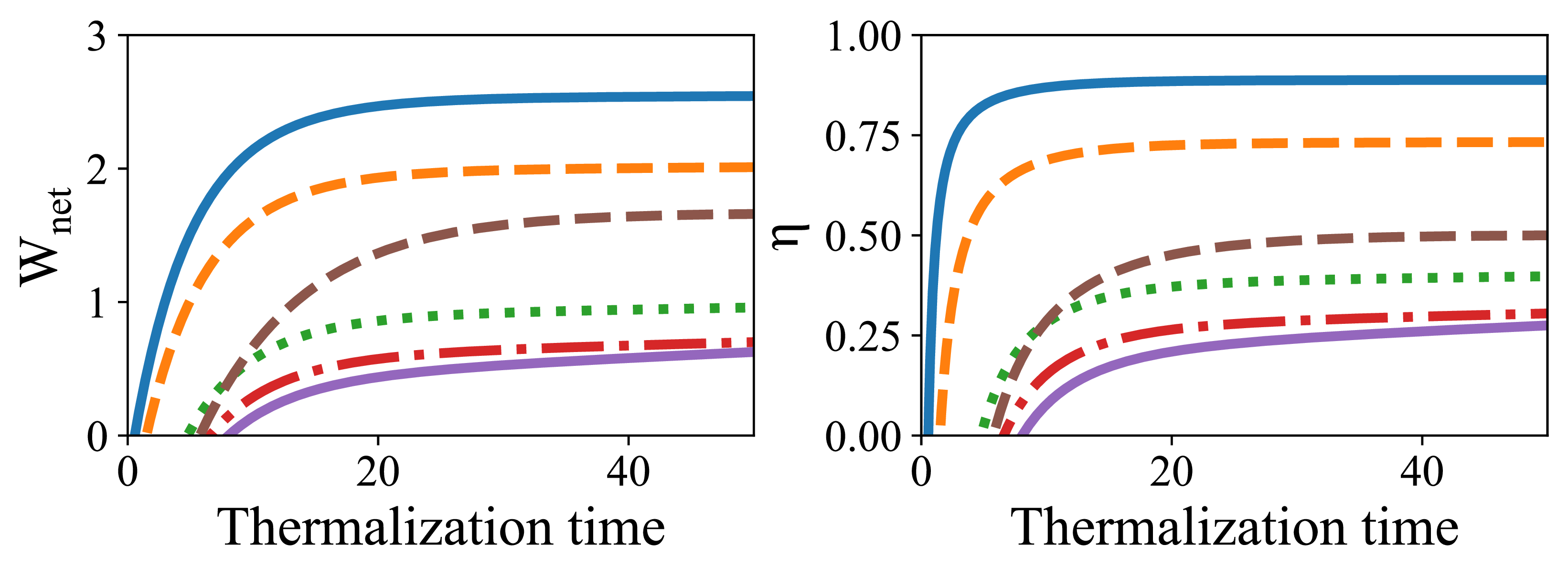}\\
\hspace{-2.75cm}{(e)\hspace{3.75cm}(f)}\\
\includegraphics[width=.5\textwidth]{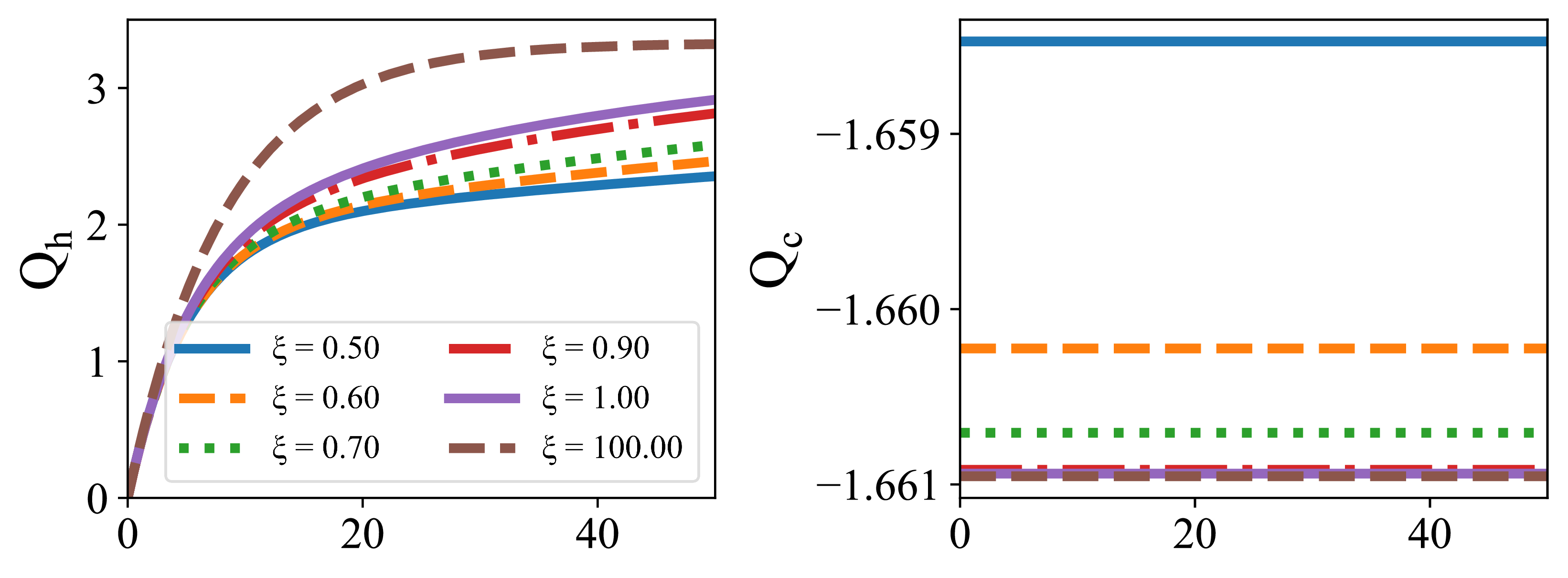}\\
\hspace{-2.75cm}{(g)\hspace{4cm}(h)}\\
\includegraphics[width=.5\textwidth]{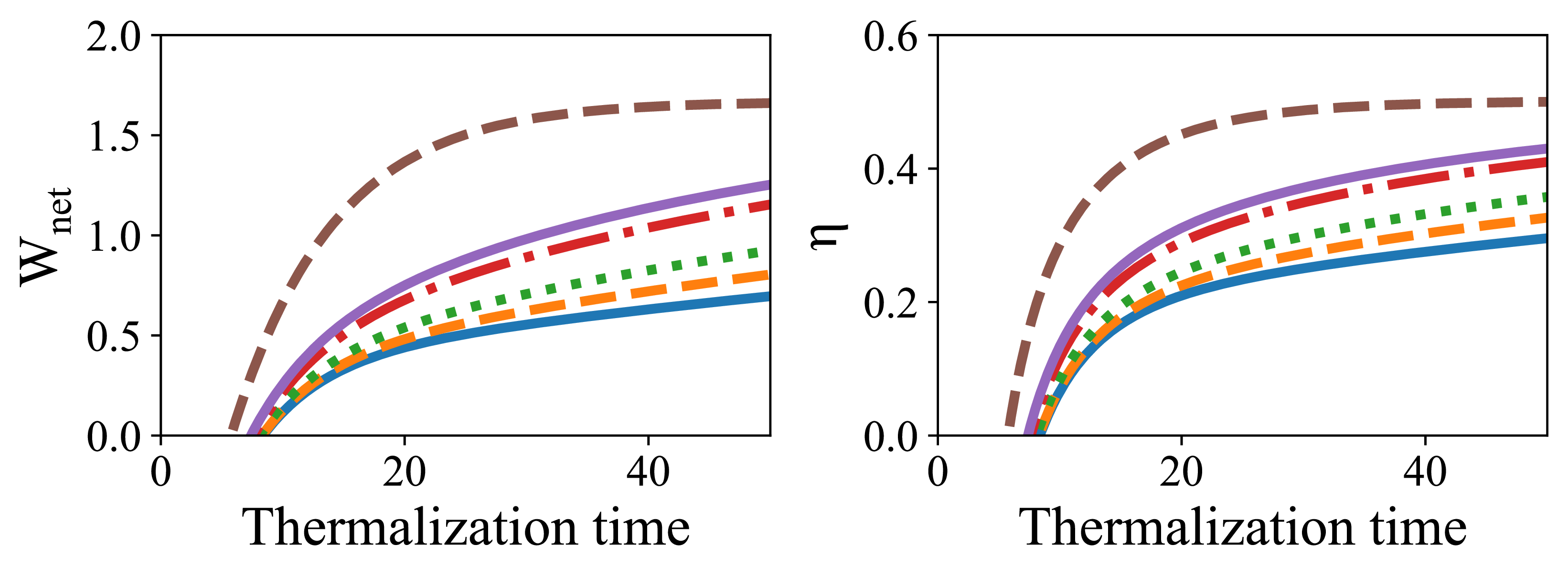}
\caption{Trapped-atom engine operation for various interatomic distances and thermalization times. Panels (a) to (d) depict the heat absorbed from the hot reservoir, heat released to the cold reservoir, total work done by the engine, and operating efficiency, respectively, for different relative interatomic distances $\xi=0.19$ (solid-blue curve), $\xi=0.2$ (dashed-yellow curve), $\xi=0.25$ (dotted-green curve), $\xi=0.3$ (dash-dotted red curve), $\xi=0.4$ (solid-indigo curve), and $\xi=100$ (dashed-maroon curve). Panels (e) to (h) display the corresponding results for $\xi=0.5, 0.6, 0.7, 0.9, 1$, and $100$. The rest parameters are the same as in Fig. \ref{f2}.}\label{f3}
\end{center}
\end{figure}

Moreover, the influences of photon-induced dipole-dipole interactions have become more prominent when the system is fully thermalized by the hot reservoir. Here the engine has already absorbed enough amount of heat and operates with an optimum efficiency solely determined by light-induced DDIs without any competitions from finite-time processes. Hence DDI effects are manifested as pronounced oscillations in the thermodynamic quantities. Critical scrutiny reveals collective dissipation, which represents the combined effect of energy loss and decay within the system, is the fundamental source of these oscillations (see Appendix \ref{A} for the details). Specifically, when collective dissipation is vanished, the efficiency of the engine will become maximal. This is due to the fact that zero dissipation indicates less energy loss, which allows a greater part of the absorbed heat to be transformed into useful work. Conversely, when collective dissipation reaches its maximum positive or negative value, efficiency falls to its minimum. In such instances, high energy loss hinders the conversion of absorbed heat into work output, resulting in decreased efficiency. As a result, the oscillatory behavior of thermodynamic quantities is caused by collective dissipation. 

On the other hand, within the range of interatomic distance $\xi\in[0.176,1]$, these oscillations are not observed since collective dissipation remains consistently positive without changing sign. Interestingly, in this parameter regime, we find that the frequency shift can counteract collective dissipations, thereby promoting efficiency and enhancing work output. Nevertheless, the effectiveness of the frequency shift diminishes rapidly as the interatomic spacing falls within the range $\xi\in[0.176,1]$, where the dissipations remains almost maximal. Despite this limitation, the frequency shift still plays a crucial role in enhancing the overall system performance, particularly when $\xi\in[0.176,0.23]$, as near-perfect engine efficiency can be achieved in this range (see Appendix \ref{A} for more details). This range of interatomic distances correspond to the ideal trapped-ion engine discussed in \cite{ar8}.

The competition between light-induced DDIs and finite-time thermalization is clearly demonstrated in Figs. \ref{f3}(a-h), where we compare the behavior of strongly interacting atoms with $\xi$ as small as $0.19$ and nearly non-interacting atoms ($\xi=100$). For shorter interatomic distances, the interacting working medium efficiently absorbs heat from the hot reservoir and generates work output with higher efficiency compared to both the quasi-static and non-interacting cases \cite{ar8,ar40,ar29,ar37,ar39}. However, the advantage of collective effects from light-induced DDIs is limited to short interatomic distances. As we further increase $\xi$, the system absorbs less energy from the hot reservoir and dissipates more energy to the cold bath, as shown in Figs. \ref{f3}(a-c). Consequently, the efficiency of the engine, as depicted in Fig. \ref{f3}(d), is reduced compared to the non-interacting regime observed at $\xi=100$. In this scenario, non-interacting working media \cite{ar8,ar40}, outperform their interacting counterparts particularly for $\xi\in[0.3,1]$ as can be seen in Figs. \ref{f3}(e-h). It should also be noted that the heat lost to the cold bath remains unaffected by the thermalization time, as the projection measurement is assumed to be instantaneous \cite{ar16,ar37,ar38}. 

\begin{figure}
    \centering
    \hspace{-2.75cm}(a)\\
    \includegraphics[width=.25\textwidth]{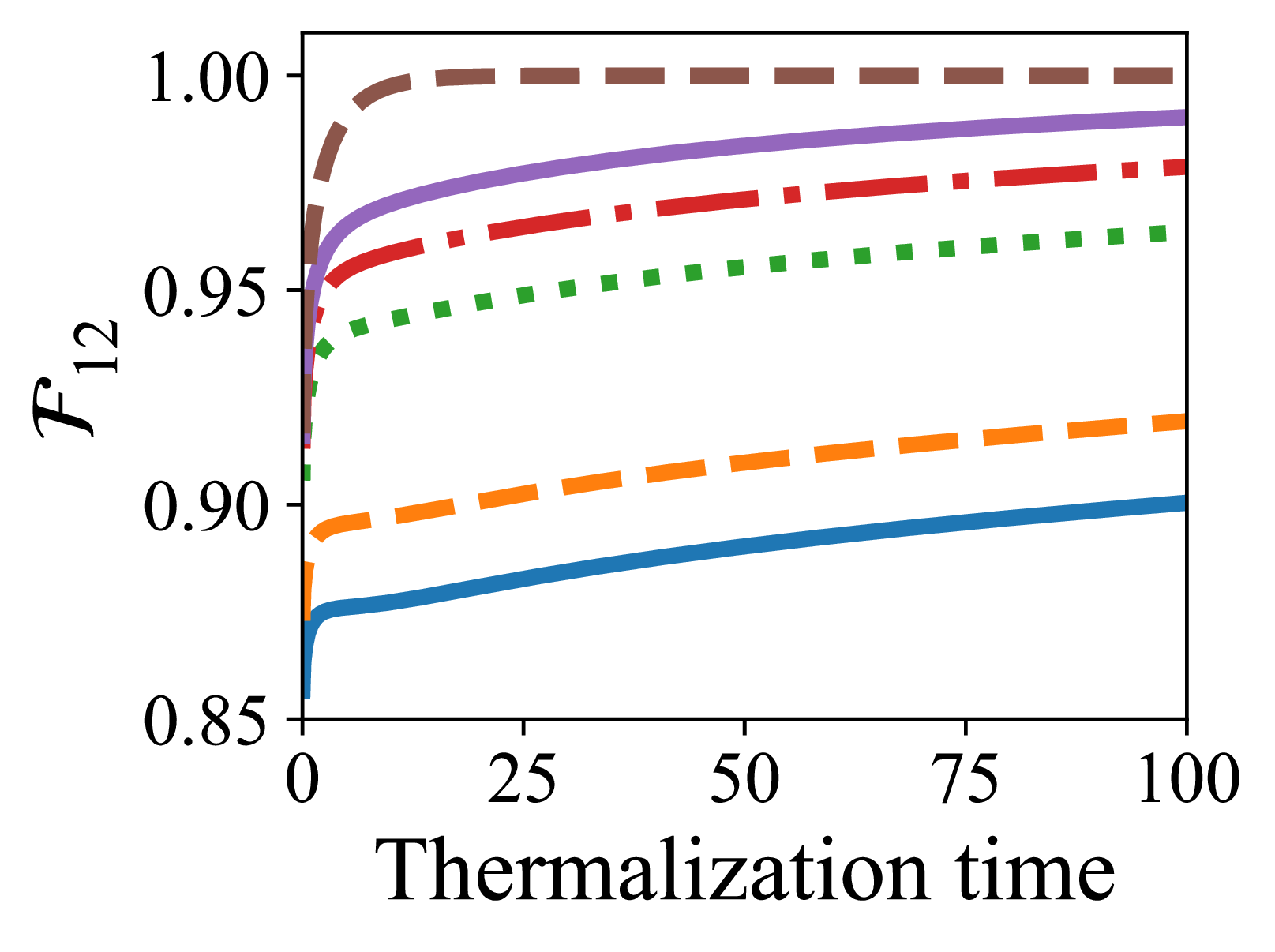}\\
    \hspace{-2.25cm}(b)\\
    \includegraphics[width=.25\textwidth]{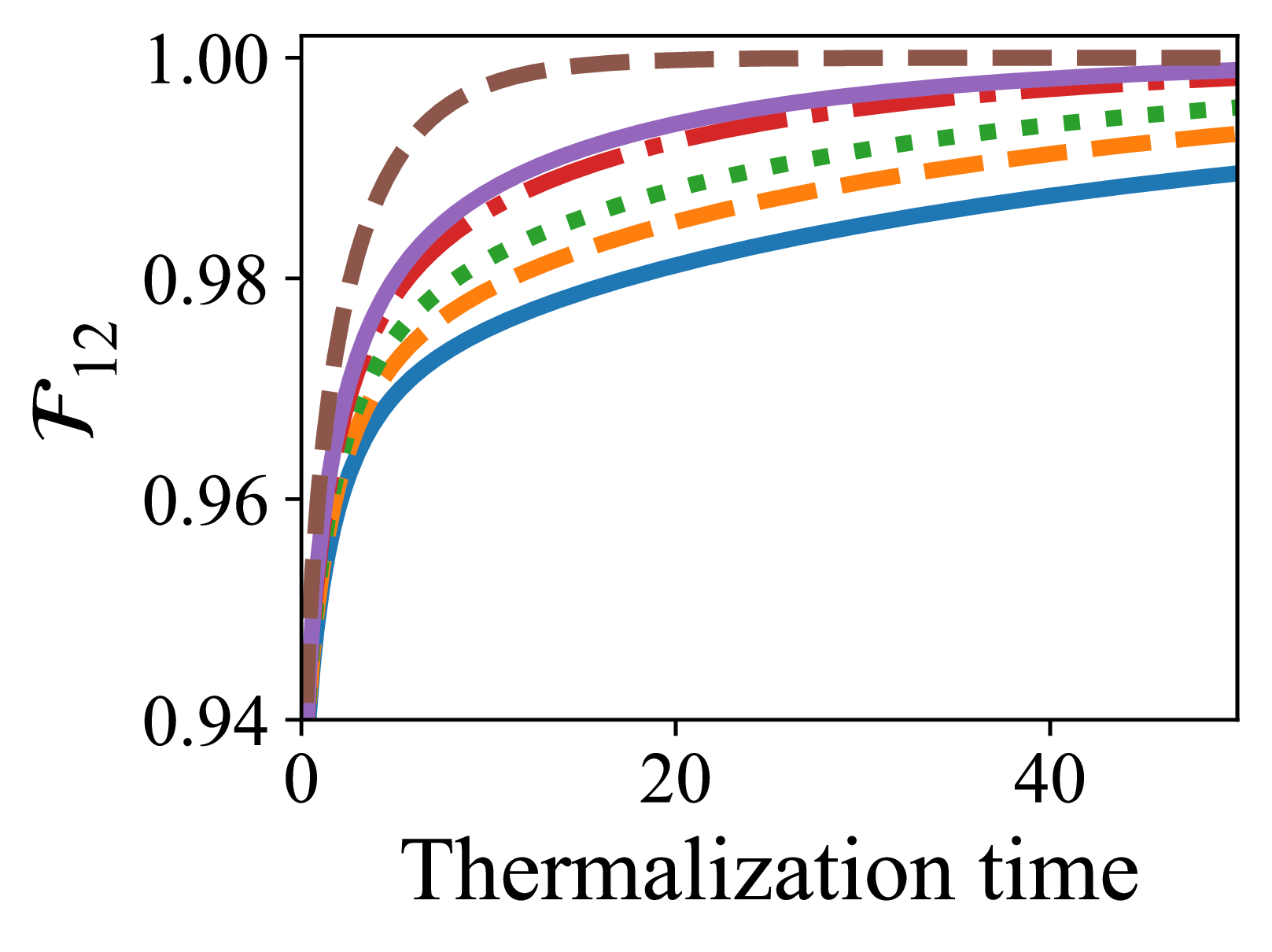}
    \caption{Fidelity of the first stroke for various interatomic distances and thermalization times. (a) Investigation of thermalization process using fidelity for different relative interatomic distances $\xi=0.19$ (solid-blue curve), $\xi=0.2$ (dashed-yellow curve), $\xi=0.25$ (dotted-green curve), $\xi=0.3$ (dash-dotted red curve), $\xi=0.4$ (solid-indigo curve), and $\xi=100$ (dashed-maroon curve) (a). (b) The corresponding results for $\xi=0.5, 0.6, 0.7, 0.9, 1$, and $100$. The rest parameters are the same as in Fig. \ref{f2}.}\label{f100}
\end{figure}

To gain further insight into thermalization process of the working medium, we plot fidelity of the first stroke as shown in Figs. \ref{f100}(a) and \ref{f100}(b) for various interatomic distances. The working medium deviates from equilibrium within the given thermalization time when interatomic distances are short, but tends to thermalize faster as interatomic distance increases. Furthermore, we show in Appendix \ref{A} that the thermodynamic quantities converge to their quasi-static counterpart when all coherences, including those induced by DDIs, are effectively washed out through a prolonged thermalization process. 
\subsection{Full thermalization and finite unitary dynamics}
\begin{figure}
    \centering
    \hspace{-2.75cm}(a)\hspace{4cm}(b)\\
    \includegraphics[width=.5\textwidth]{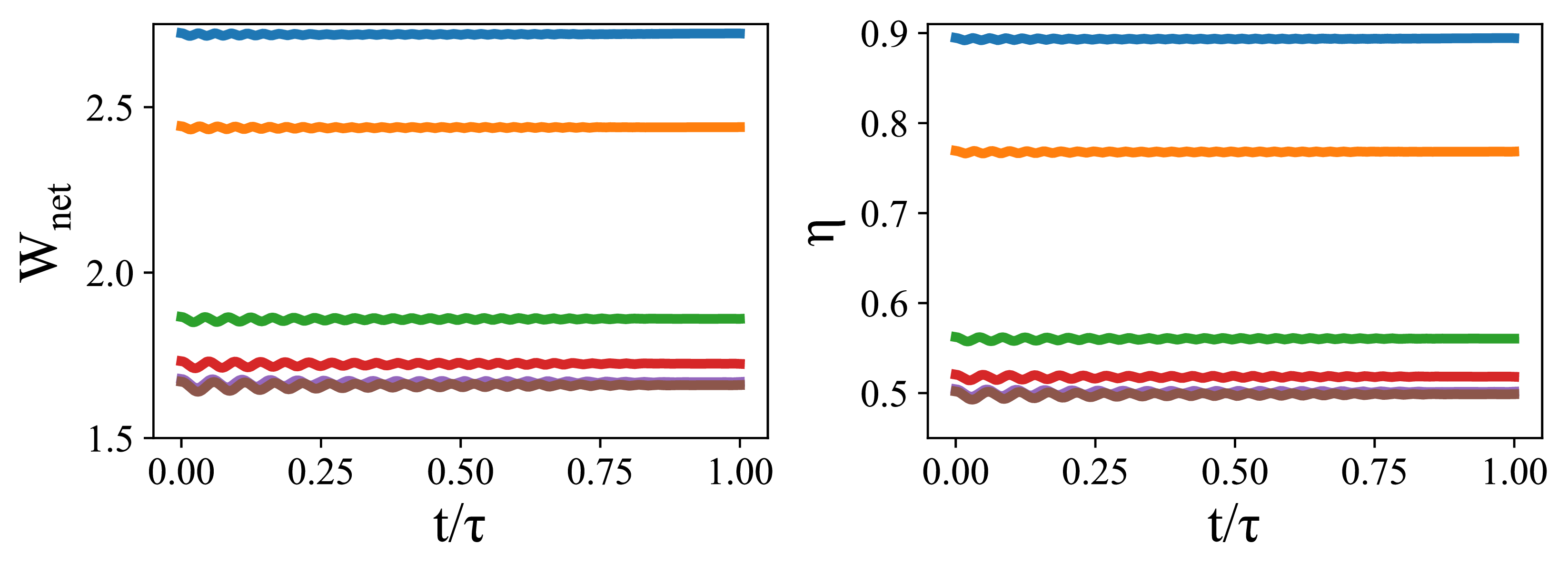}\\
    \hspace{-2.75cm}(c)\hspace{4cm}(d)\\
    \includegraphics[width=.5\textwidth]{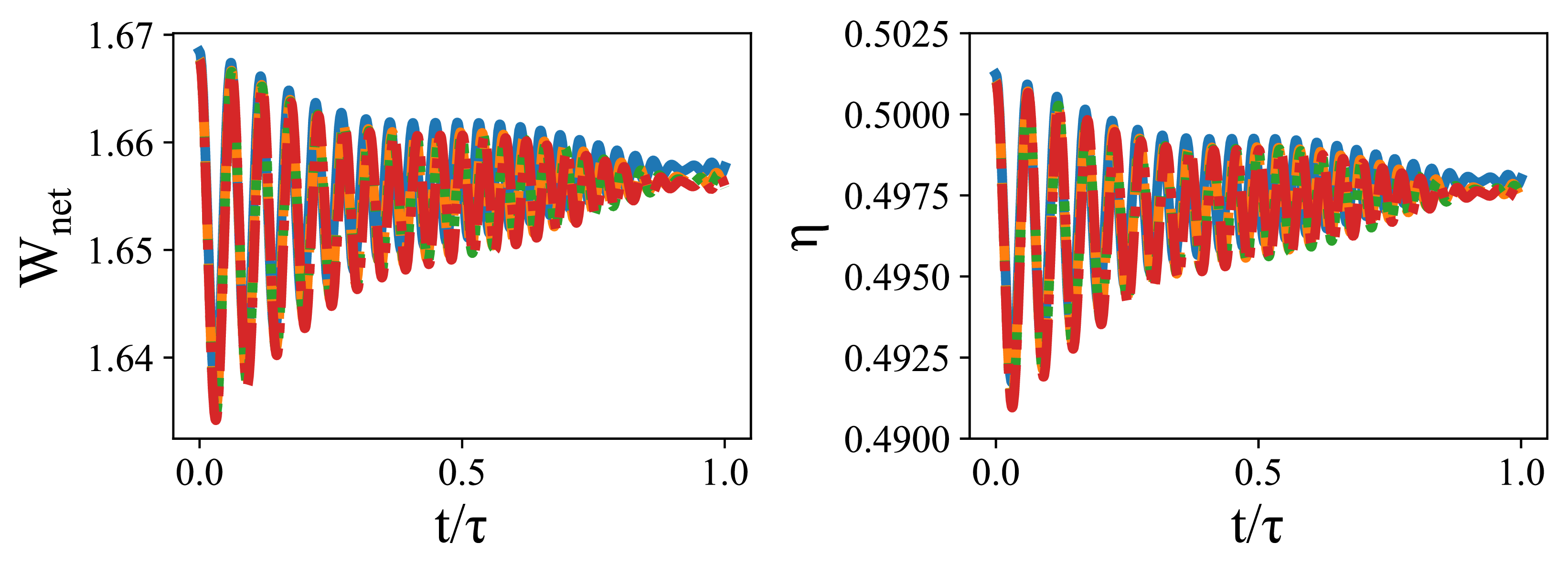}
    \caption{Trapped-atom engine operation for various interatomic distances and unitary driving times. Total work done by the engine and operating efficiency are, respectively, shown in (a) and (b) for different relative interatomic distances $\xi=0.19$ (solid-blue curve), $\xi=0.2$ (solid-yellow curve), $\xi=0.25$ (solid-green curve), $\xi=0.3$ (solid red curve), $\xi=0.4$ (solid-indigo curve), and $\xi=0.5$ (solid-maroon curve). The last two curves overlap each other. The corresponding results for $\xi=0.6, 0.8, 1$, and $100$ are shown in (c) and (d). Beyond, $\xi>0.5$, these results are barely distiguishable. The unitary evolution lasts for $\tau=10/\omega$, and the rest parameters are the same as in Fig. \ref{f2}.}\label{f4}
\end{figure}
In this section, we analyze the operation of the engine assuming that the working medium is fully thermalized by the hot reservoir, while a time dependent magnetic field drives the system within a finite-time interval \cite{ar1,ar38}. Finite-unitary driving excites the working medium, leading to the generation of quantum friction, which in turn affects the performance of the engine \cite{ar5,ar36}. This time-dependent driving is a distinctive characteristic of genuine quantum engines and has attracted numerous attention in the field of finite-time quantum thermodynamics \cite{ar1,ar2,ar5,ar26,ar50}. In Figs. \ref{f4}(a) and \ref{f4}(b), we present that the engine generates a significant work output with high efficiency for short interatomic distances. Specifically, we observe an efficiency of approximately $90\%$ for $\xi=0.19$, $77.5\%$ for $\xi=0.2$, and $56\%$ for $\xi=0.25$. We also notice minor improvements in efficiency for interatomic distances $\xi\geq0.3$. This trend is further illustrated in Figs. \ref{f4}(c) and \ref{f4}(d), where the work output and efficiency exhibit slight variations compared to the previous cases with short interatomic distances. Therefore, when the system undergoes unitary evolution and has already absorbed sufficient energy from the hot reservoir, the performance of the engine is enhanced through a collective frequency shift. This improvement occurs even when there is coherence generated during the finite-unitary driving process. 

The efficiency achieved at the sudden unitary limit ($t/\tau\rightarrow 0$) is due to the projection measurement protocol implemented in the third stroke \cite{ar38}. This protocol sets the initial state of the fourth stroke before the build-up of coherence effects in the unitary compression step (4-1). Consequently, the engine operates with remarkable efficiency at the sudden limit, benefiting from the instantaneous nature of the projection measurement. Moreover, for short interatomic distances, the frequency shift overcomes coherence effects, leading to negligible oscillations of thermodynamic quantities, as demonstrated in Figs. \ref{f4}(a) and \ref{f4}(b). These oscillations arise from the combined effects of the driving protocols and the overlap of states of the working medium with the projected state. On the other hand, when interatomic distances are large, the frequency shift becomes insignificant, resulting in slightly damped oscillations caused by coherences generated in the second and fourth strokes (see Figs. \ref{f4}(c) and \ref{f4}(d)). Eventually, these oscillations vanish over a relatively long unitary driving period. These findings are particularly relevant to systems with short interatomic spacing, such as nitrogen-vacancy color centers and other solid-state systems \cite{ar59}, and superconducting qubits \cite{ar58}. They provide valuable insights and serve as a stepping stone for experimentalists working in these areas. In the following section, we further investigate the effect of many-body coherence on the engine operation by considering both finite-time thermalization and unitary driving.
\subsection{Both finite thermalization and unitary dynamics}
\begin{figure}
    \centering
    \hspace{-2.75cm}(a)\hspace{4cm}(b)\\
    \includegraphics[width=.5\textwidth]{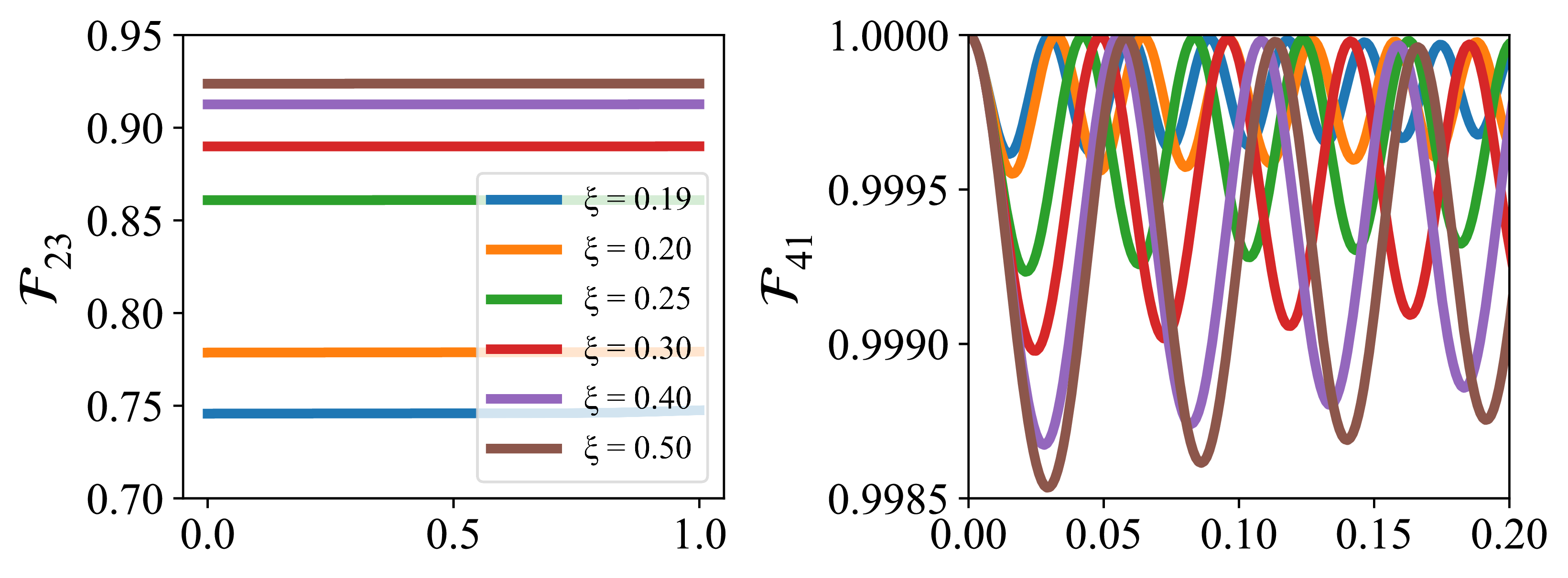}\\
    \hspace{-2.75cm}(c)\hspace{4cm}(d)\\
    \includegraphics[width=.5\textwidth]{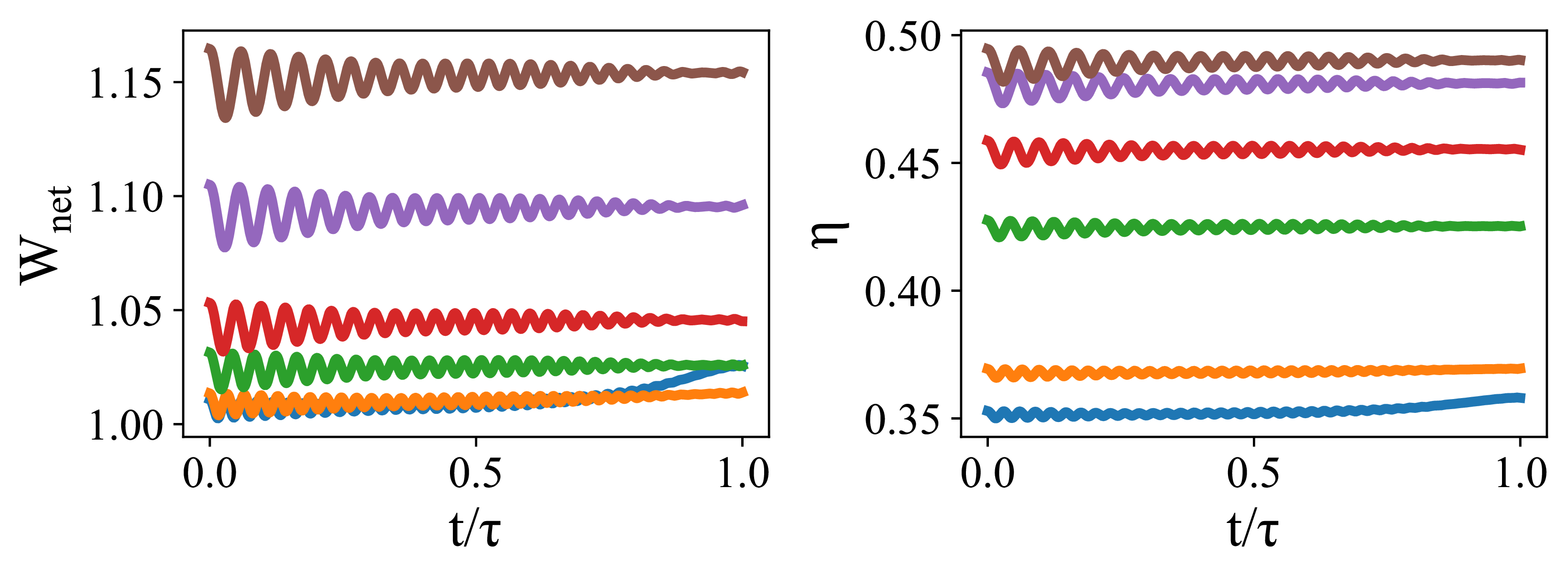}\\
    \hspace{-2.75cm}(e)\hspace{4cm}(f)\\
    \includegraphics[width=.5\textwidth]{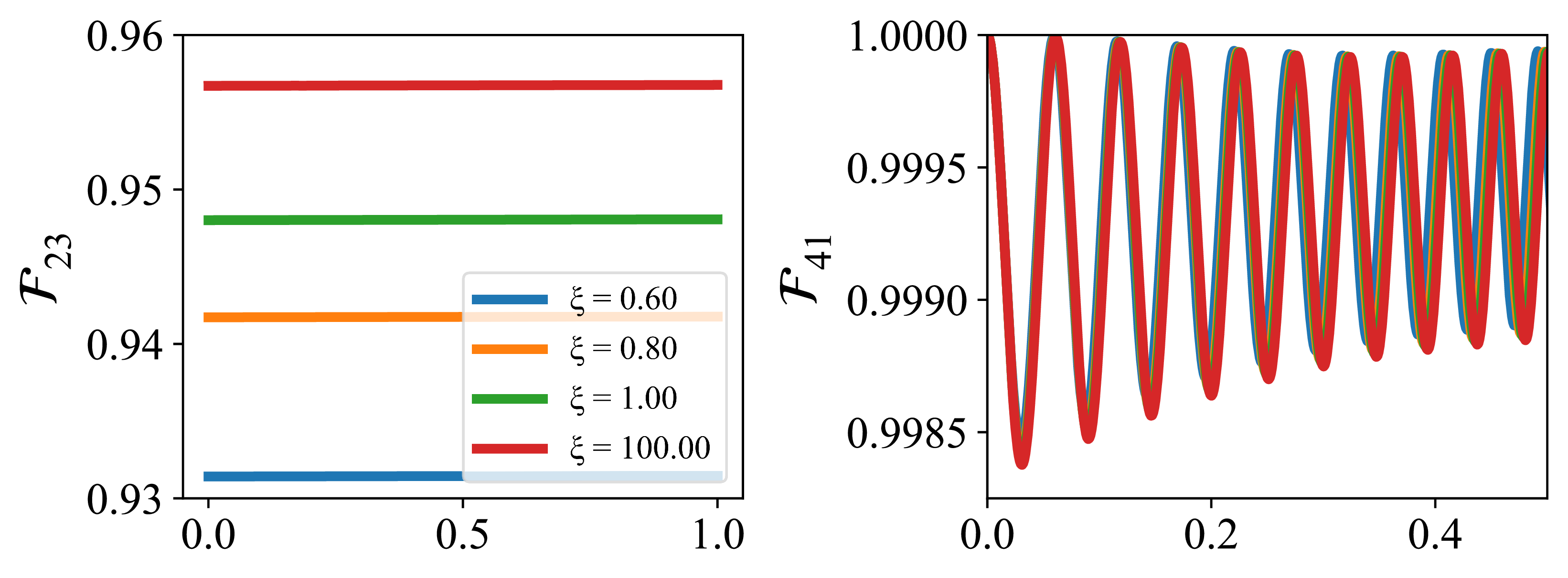}\\
    \hspace{-2.75cm}(g)\hspace{4cm}(h)\\
    \includegraphics[width=.5\textwidth]{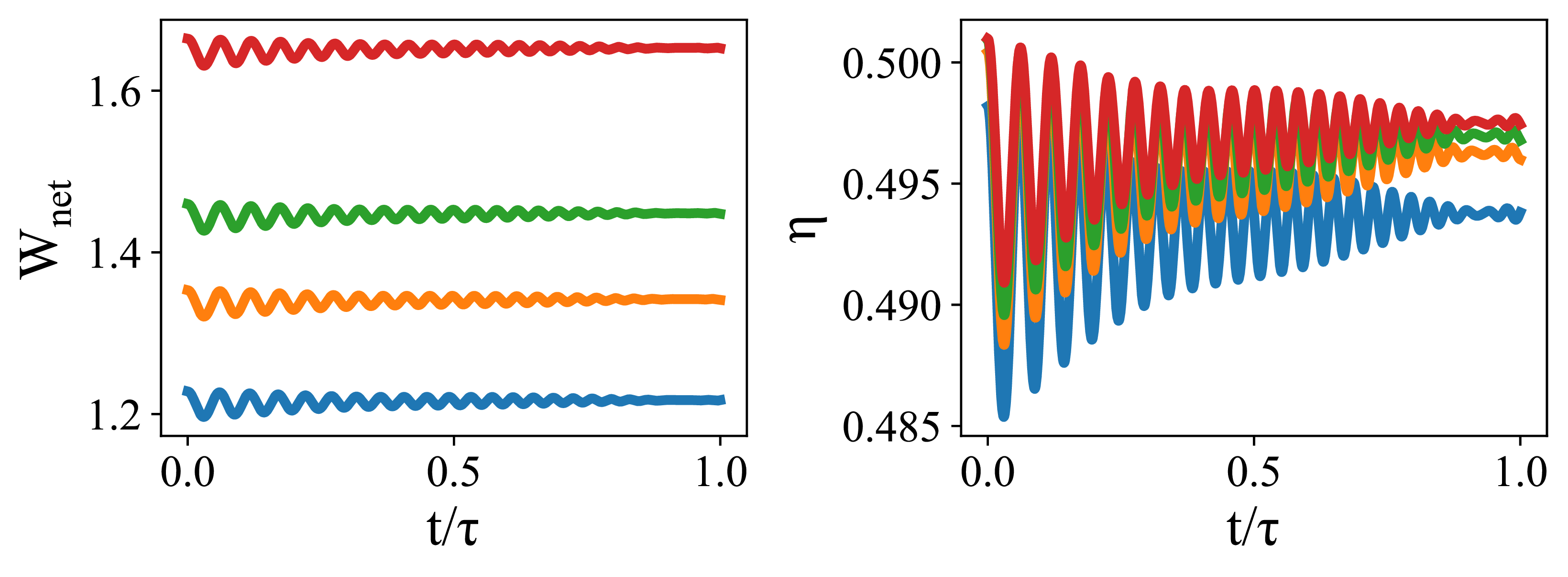}
    \caption{Trapped-atom engine operation for various interatomic distances and unitary driving times. Fidelity of second and forth strokes, total work done by the engine, and operating efficiency are, respectively, depicted from (a) to (d) for different relative interatomic distances $\xi=0.19$ (solid-blue curve), $\xi=0.2$ (solid-yellow curve), $\xi=0.25$ (solid-green curve), $\xi=0.3$ (solid red curve), $\xi=0.4$ (solid-indigo curve), and $\xi=0.5$ (solid-maroon curve). Panels (e) to (h) display the corresponding results for $\xi=0.6, 0.8, 1$, and $100$. The thermalization process and unitary evolution last for $50/\omega$ and $10/\omega$, respectively, and the rest parameters are the same as in Fig. \ref{f2}.}\label{f5}
\end{figure}
Finally, we investigate the effects of both finite thermalization and finite unitary driving conditions. We set the thermalization time to $50/\omega$, while the unitary driving occurs for $\tau=10/\omega$, as discussed in the previous section. The presence of quantum coherence, which is not completely erased by the hot reservoir, impacts the unitary driving protocols that generate real-time work output, efficiency, and power, as depicted in Figs. \ref{f5}(a-h). Moreover, this finite-time operation yields a finite-power output defined as ${\cal P}_{output}=(W_{23}+W_{41})/t_{cycle},$ where $t_{cycle}$ is the total time taken throughout the engine operations.

In Figs. \ref{f5}(a) and \ref{f5}(b), we analyze the fidelity \cite{ar1} of the second and fourth strokes to investigate the irreversibility of the engine cycle caused by finite-time operation \cite{ar5,ar29}. The result shown in Fig. \ref{f5}(a) represents the distance between the time-dependent state of the working medium and the thermal equilibrium condition that would be achieved if the unitary driving were performed quasi-statically throughout the expansion process. We observe that for short inter-atomic distances, the system deviates significantly from the equilibrium condition, resulting in significant quantum friction that leads to a decrease in work output and efficiency, as elucidated below. 

Furthermore, we observe excitations of the system during the fourth stroke, as illustrated in Fig. \ref{f5}(b), even though the system is projected onto the ground state before the fourth stroke begins. It is important to note that coherence from the previous stroke is completely eliminated by this projection protocol used in the third stroke and using the projection state as the initial state for the fourth stroke. Therefore, the excitation observed in the fourth stroke is solely due to finite unitary driving. We find that closely spaced atoms exhibit a weaker response to the finite unitary driving compared to nearly non-interacting atoms.

The significantly reduced work output and efficiency observed in Figs. \ref{f2}(c) and \ref{f2}(d), in comparison with the results shown in Fig. \ref{f4}, are primarily attributed to the residual coherence that persists during the finite thermalization time and is subsequently transferred to the second unitary stroke. On the other hand, Figs. \ref{f2}(g) and \ref{f2}(h) show the residual coherence has small effect on the weakly interacting working medium due to the rapid thermalization caused by the hot reservoir (see Figs. \ref{f100}(a) and \ref{f100}(b)). These results indicate strongly interacting working medium generate strong quantum friction during finite-time, resulting in reduced performance of the quantum heat engine. The effect of quantum friction slightly decreases in a relatively long unitary driving time since in this case the system moves towards the quasi-static condition \cite{ar37,ar29}, and this point is further illustrated in Appendix \ref{B} (see Fig. \ref{f12}).

\begin{figure}
    \begin{center}
    \hspace{-3.75cm}(a)\\
    \includegraphics[width=.25\textwidth]{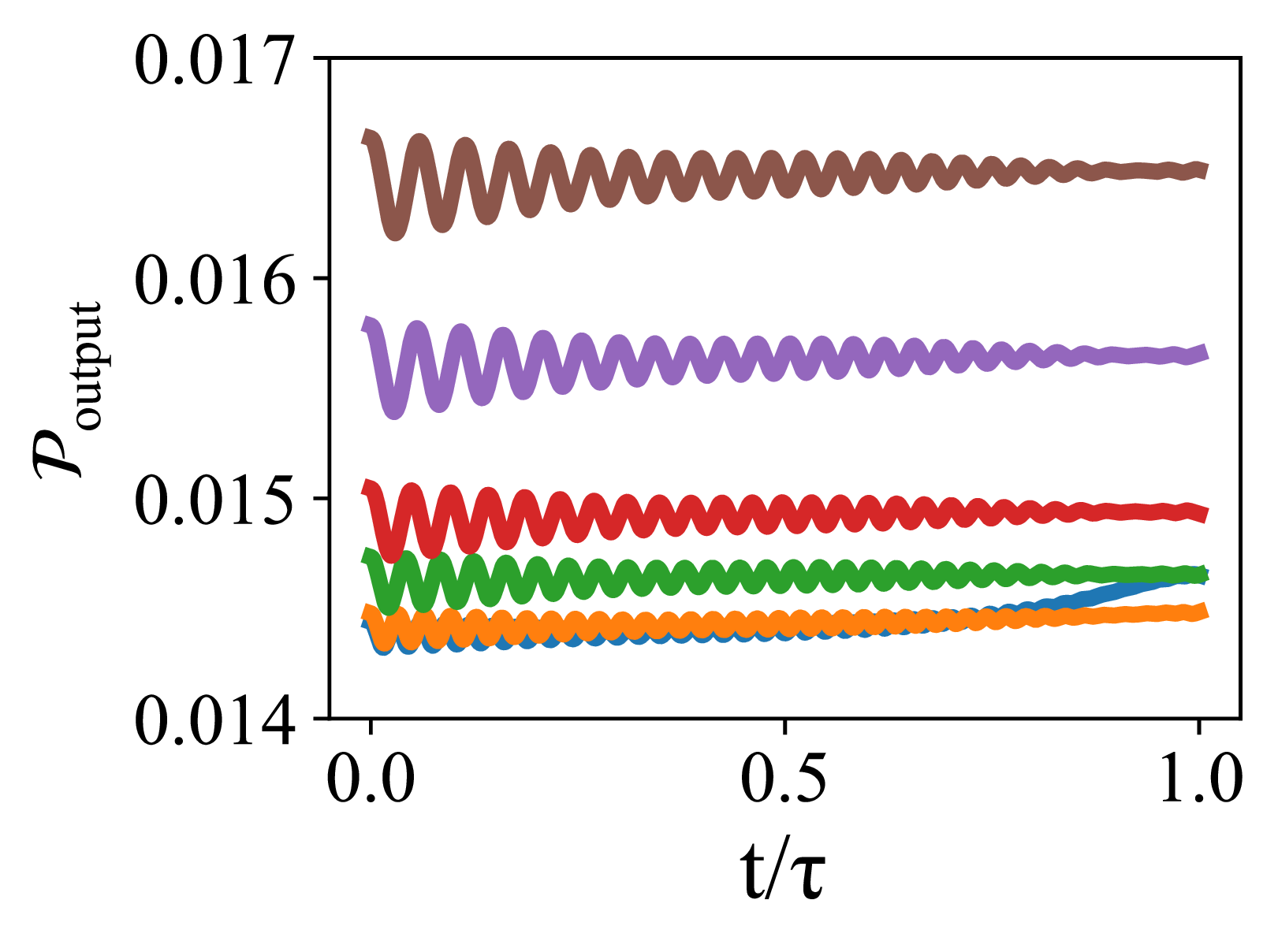}\\
    \hspace{-3.75cm}(b)\\
    \includegraphics[width=.25\textwidth]{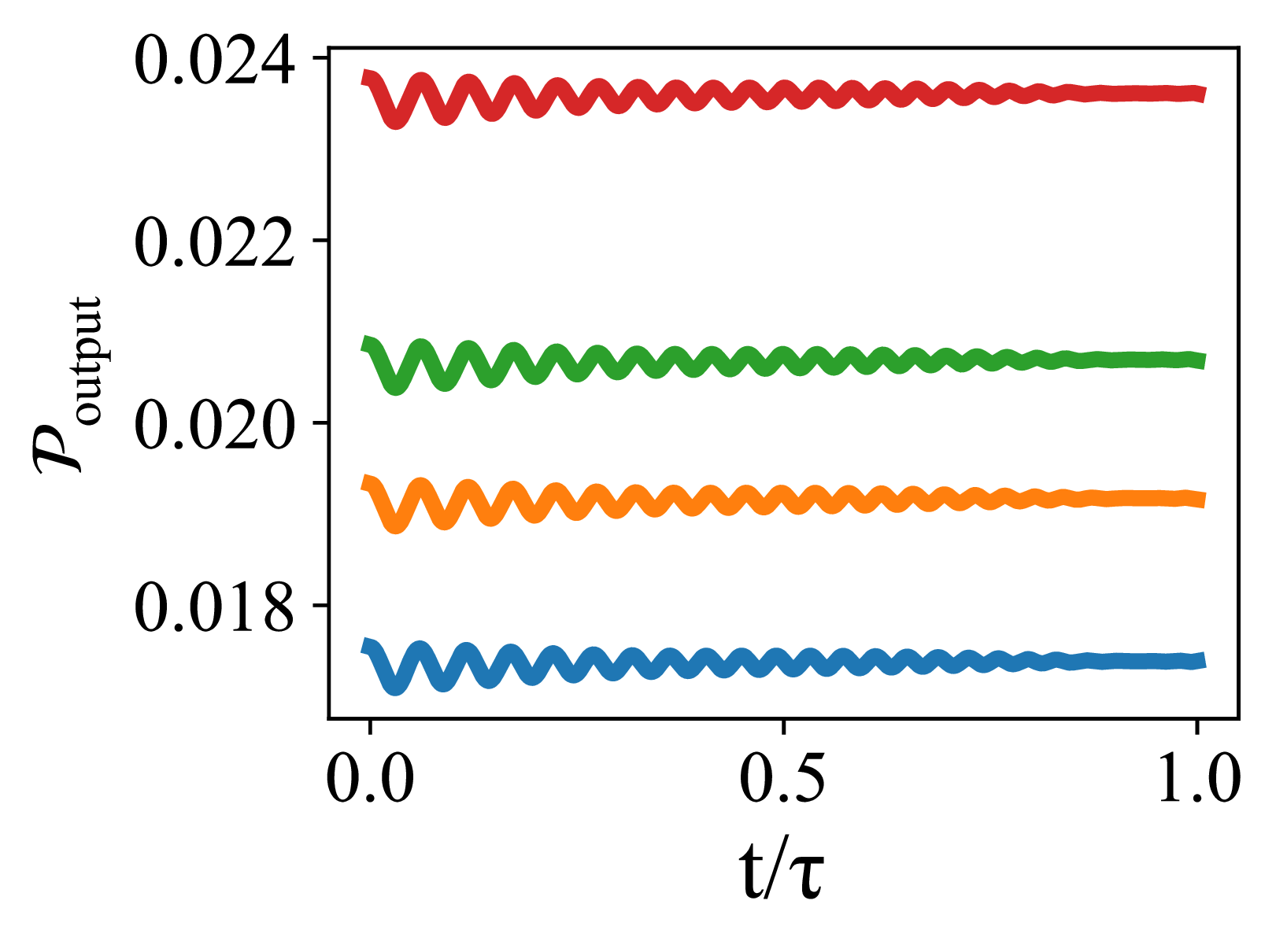}
    \caption{Power output of trapped-atom engine for various inter-atomic distances and unitary driving times. (a) Power output for different relative inter-atomic distances $\xi=0.19$ (solid-blue curve), $\xi=0.2$ (solid-yellow curve), $\xi=0.25$ (solid-green curve), $\xi=0.3$ (solid red curve), $\xi=0.4$ (solid-indigo curve), and $\xi=0.5$ (solid-maroon curve). (b) The thermalization process and unitary evolution last for $50/\omega$ and $10/\omega$, respectively, and the rest parameters are the same as in Fig. \ref{f2}.}\label{f6}
    \end{center}
\end{figure}

Finite-time consideration in quantum heat engine is intriguing since it leads to finite power output, as can be seen in Figs. \ref{f6}(a) and \ref{f6}(b). In the case of a quantum Otto engine with an interacting many-body working medium, the finite-time operation degrades performance when compared to non-interacting cases. However, the quantum world offers alternatives, such as projection measurement, which can eliminate the requirement for conventional thermodynamic reservoirs and coherences depending on the measurement basis used. To further improve power output and efficiency, the shortcut to adiabaticity technique proposed in \cite{ar36} and implemented in \cite{ar64,ar65} can also be used to mitigate the adverse effects of quantum coherence in quantum engine operation. In addition, a quick route to thermalization has been proposed in \cite{ar66,ar67} to boost the power output of quantum heat engines. Combining these strategies, the trapped-atom Otto engine under consideration would be a feasible option with the potential to achieve both finite power output and enhanced engine efficiency, effectively overcoming the trade-off between the two.  
\section{\label{V} conclusion}
We have investigated the finite-time operation of a quantum Otto engine using a working medium consisting of atoms trapped in the Lamb-Dicke regime \cite{ar8,ar30,ar35} and interacting via photon-mediated dipole-dipole interactions \cite{ar24,ar30}. The isochoric strokes of the engine are implemented using quantum mechanical projection techniques \cite{ar15,ar16,ar17} and characteristics of open quantum systems \cite{ar24,ar30,ar31}, while the unitary strokes are implemented by subjecting the working medium to time-dependent magnetic fields. After their projection onto the ground state, the atoms undergo a unitary compression to complete the engine cycle with a minimal energy dissipation. Alternatively, a light pulse \cite{ar53,ar54} can be applied to reset the atomic initial states per cycle, as a feedback loop \cite{ar16,ar55}. This setup provides great flexibility and independent control of finite-time performance of the engine with collective effects arising from dipole-dipole interactions and many-body coherences, which have been primarily limited to the single-particle case thus far \cite{ar9,ar29,ar37,ar39}. 

We analyze the finite-time performance of the trapped-atom Otto engine by considering three different scenarios: (1) finite thermalization time and adiabatic condition, (2) full thermalization and finite-time unitary dynamics, and (3) both finite thermalization time and unitary dynamics. In the finite thermalization and adiabatic condition, the operation of the engine is determined by the competition between collective dissipations and frequency shift, while the other parameters are fixed. For short interatomic distances, a collective frequency shift enables the engine to operate with nearly perfect efficiency by overcoming the effect of collective dissipations, which is the fundamental source of oscillations observed in thermodynamic quantities during finite thermalization and adiabatic conditions. Specifically, for relative interatomic spacing within the range $\xi\in[0.18,0.195]$, the thermalization time decreases by more than 60-fold, and engine efficiency is increased at the same time. However, for smaller $\xi$ values, the collective frequency shift abruptly distorts the atomic energy levels, rendering the system ineffective as an engine \cite{ar8}. For larger $\xi$ values, the collective frequency shift rapidly falls and is unable to counteract energy loss and decay caused by collective dissipations. Therefore, the advantage of light-induced DDIs for engine operation is limited to short interatomic distances, where the system is out of equilibrium for a long thermalization time. This non-equilibrium many-body system also exhibits other interesting effects, such as the emergence of unique and stable quantum phases and phase transitions as discussed elsewhere, for instance in Ref. \cite{ar58}. These findings suggest the feasibility of implementing a trapped-atom Otto engine in practical applications, especially for experimentalists working with systems featuring small interatomic spacing, such as nitrogen-vacancy color centers, superconducting qubits, and solid-state systems \cite{ar58,ar59}, offering potential avenues for achieving high-efficiency engine performance while mitigating coherence effects.

Moreover, when the system undergoes unitary evolution after it has already absorbed sufficient energy from the hot reservoir, the performance of the engine is enhanced through a collective frequency shift and projection measurements that allow the engine to operate efficiently in the sudden unitary limit. This improvement occurs even when there is coherence generated during the finite unitary driving process. Engine performance under both finite thermalization and unitary driving conditions is significantly hindered by the residual quantum coherence that is not completely erased by the hot reservoir during the finite thermalization process. Specifically, for short interatomic distances, the system deviates from the equilibrium condition and retains residual coherences that generate quantum friction against engine efficiency. For many-body systems with light-induced DDIs, this quantum friction is significant for short thermalization time and thus reduces engine efficiency, while long thermalization time leads to vanishing output power. This trade-off relation between efficiency and power output is linked to quantum coherence which can be reduced by quantum mechanical projection measurement or tuning appropriate driving fields.

To sum up, the interacting trapped-atom engine exhibits better performance in the finite-time operation compared to the quasi-static condition when considering either finite thermalization or finite unitary strokes at a time. However, when both finite-time thermalization and unitary driving are taken into account, a significant buildup of quantum friction occurs during the finite-time thermalization process, reducing the engine performance. Here we investigated a quantum Otto engine with an interacting many-body working medium and quantum projection technique, which allows single-reservoir engine operation and reduces quantum friction depending on the measurement basis. To further improve power output and efficiency, the shortcut to adiabaticity technique proposed in \cite{ar36} and implemented in \cite{ar64,ar65} can be applied to mitigate the adverse effects of quantum coherence in quantum engine operation. In addition, a quick route to thermalization has been proposed in \cite{ar66,ar67} to boost the power output of quantum heat engines. Combining these strategies, the trapped-atom Otto engine under consideration would be a feasible option with the potential to achieve both finite power output and enhanced engine efficiency, effectively overcoming the trade-off between the two. Moreover, engine operation in a full dephasing mode would be worth investigations since this approach can completely erase the residual coherence without cost \cite{ar29,ar37}.
\begin{acknowledgments}
We acknowledge support from the National Science and Technology Council (NSTC), Taiwan, under the Grant No. NSTC-112-2119-M-001-007, and from Academia Sinica under Grant AS-CDA-113-M04. We are also grateful for support from TG 1.2 of NCTS at NTU.
\end{acknowledgments}
\appendix
\section{\label{A}Finite thermalization and adiabatic condition}
\begin{figure}
\begin{center}
\hspace{-2.5cm}{(a)\hspace{3.5cm}(b)}\\
\includegraphics[width=.45\textwidth]{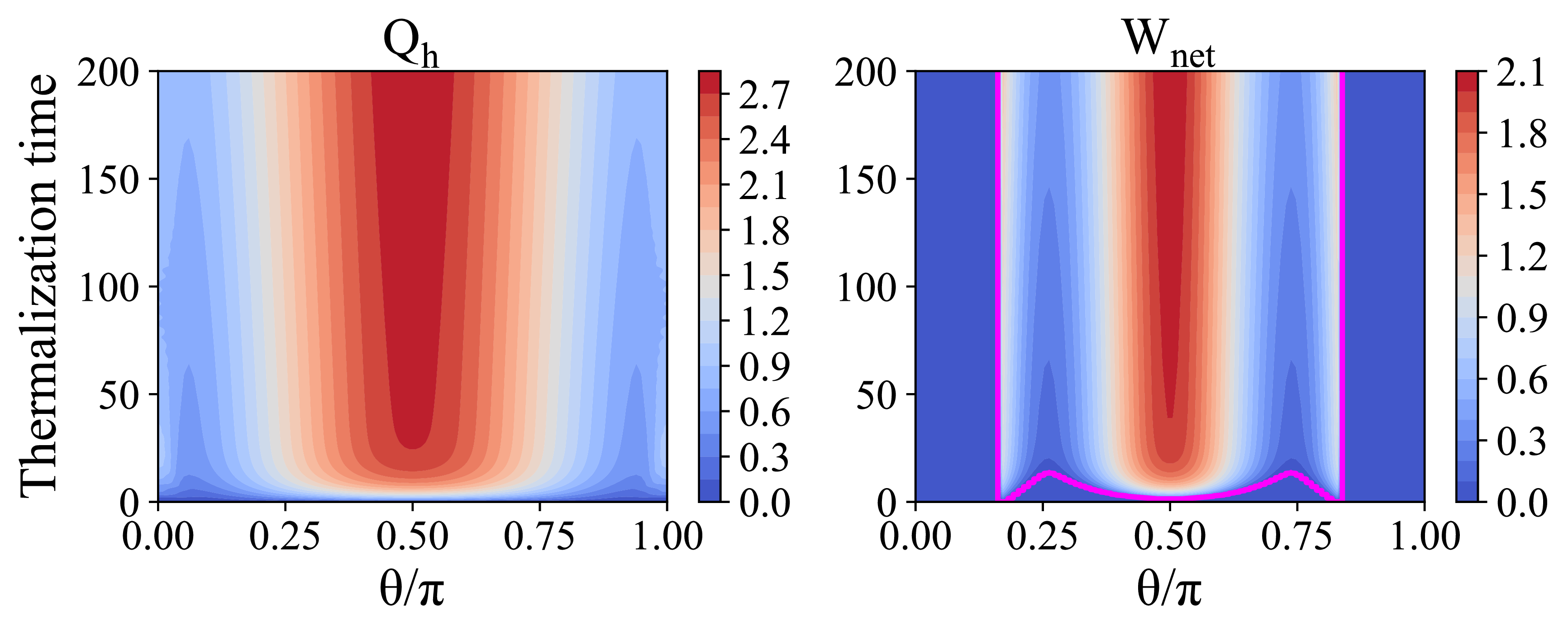}\\
\hspace{-2.5cm}{(c)\hspace{3.5cm}(d)}\\
\includegraphics[width=.45\textwidth]{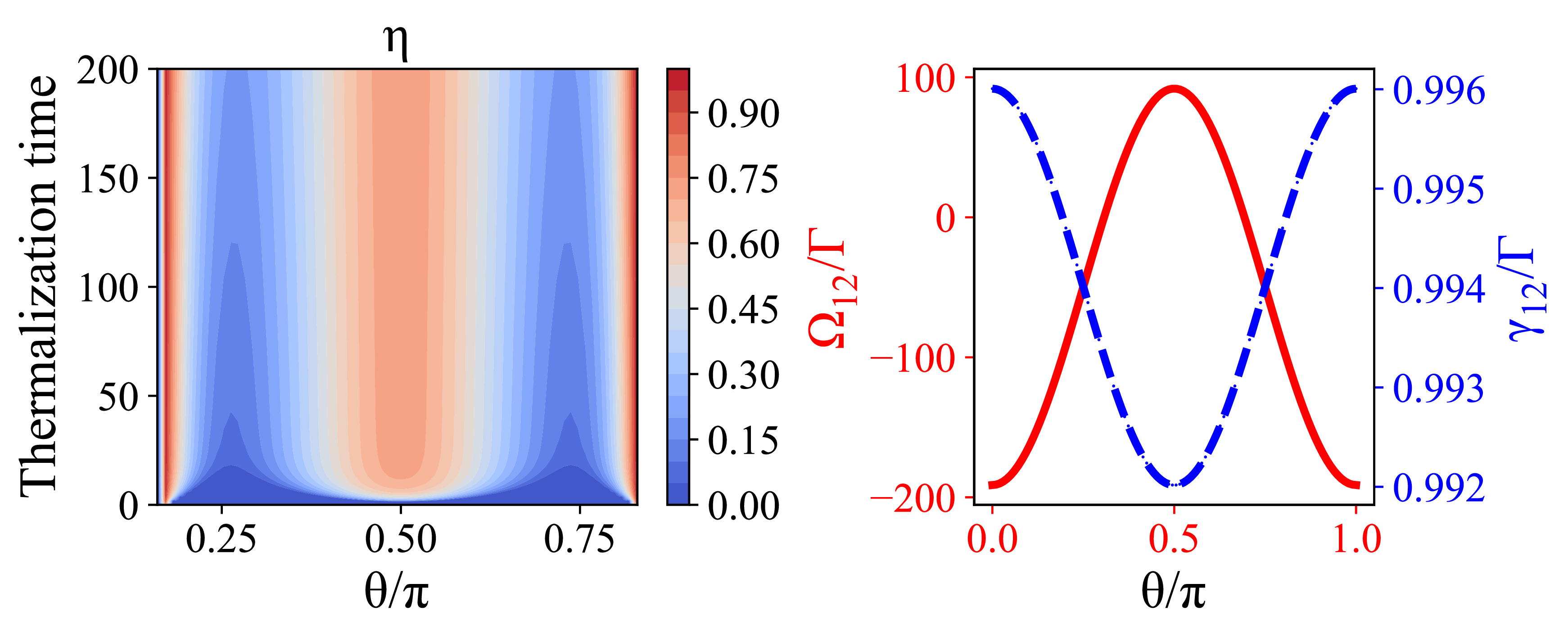}\\
 \caption{Trapped-atom engine operation for different dipole orientations ($\theta\in[0,\pi]$) and thermalization times. Heat absorbed from the hot reservoir, total work done by the engine, operating efficiency, and light-induced DDIs are respectively shown from (a) to (d) for $\xi=0.2$. The other parameters are the same as in Fig. \ref{f2}.}\label{f8} 
\end{center}
\end{figure}

\begin{figure}
\begin{center}
\hspace{-2.5cm}{(a)\hspace{3.5cm}(b)}\\
\includegraphics[width=.4\textwidth]{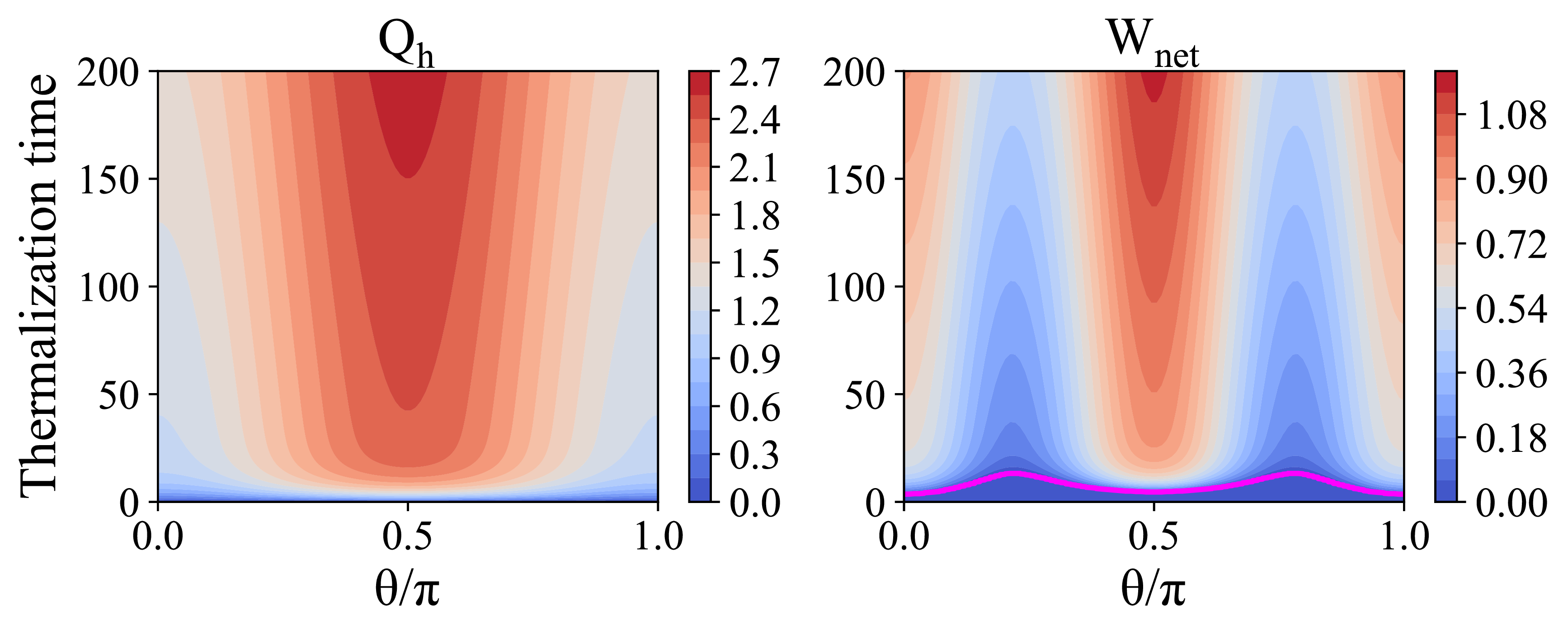}\\
\hspace{-2.5cm}{(c)\hspace{3.5cm}(d)}\\
\includegraphics[width=.4\textwidth]{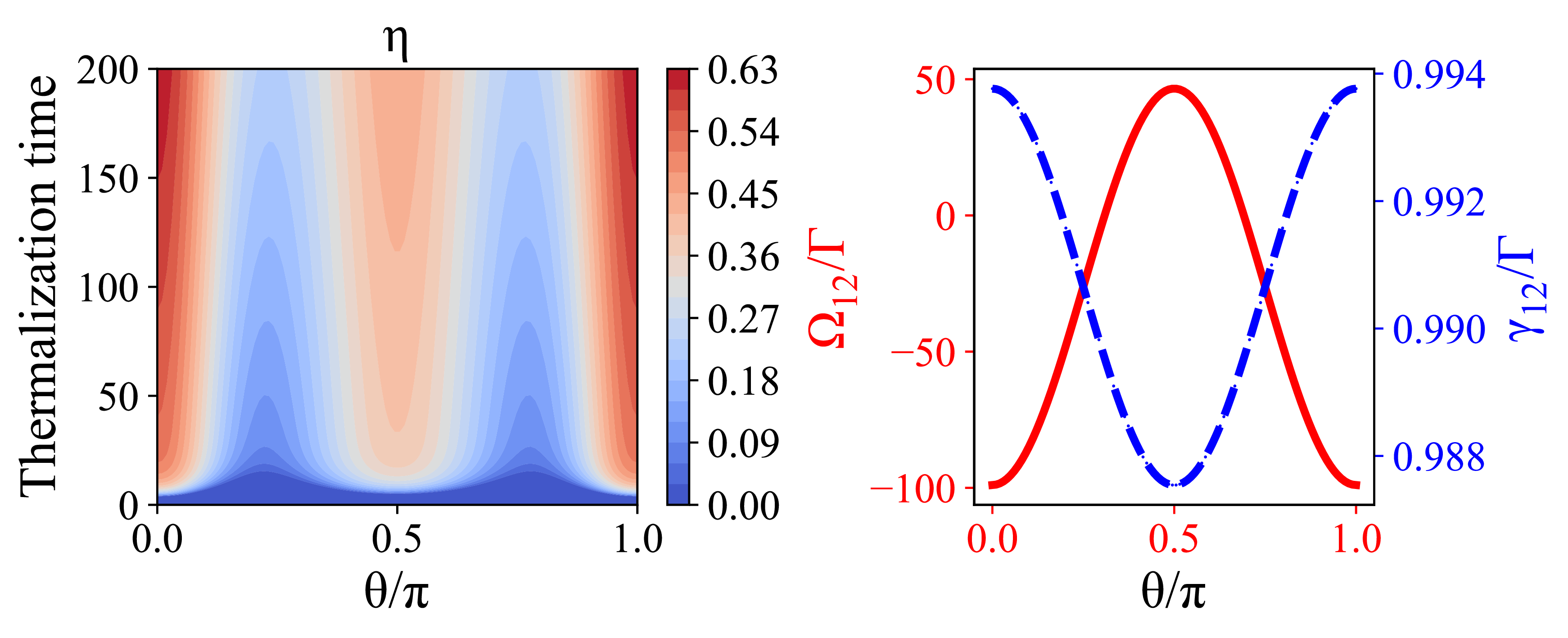}\\
\hspace{-2.5cm}{(e)\hspace{3.5cm}(f)}\\
\includegraphics[width=.4\textwidth]{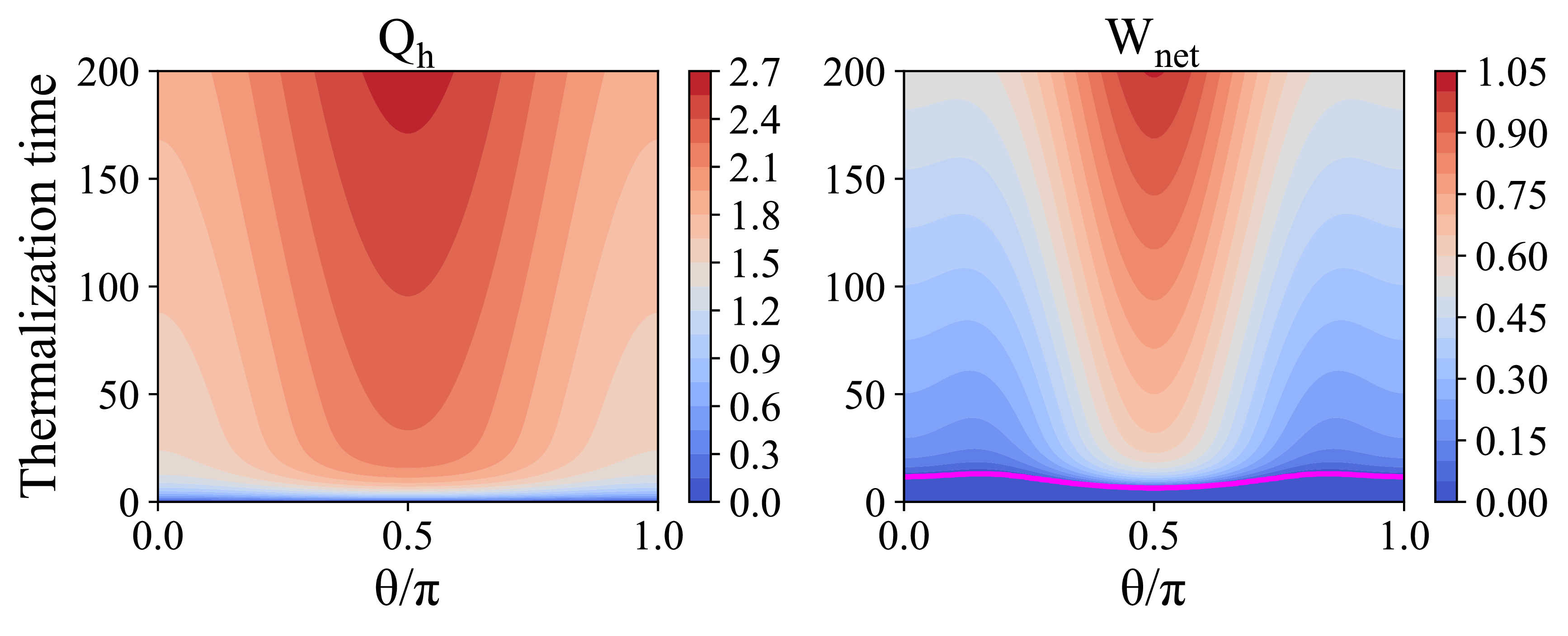}\\
\hspace{-2.5cm}{(g)\hspace{3.5cm}(h)}\\
\includegraphics[width=.4\textwidth]{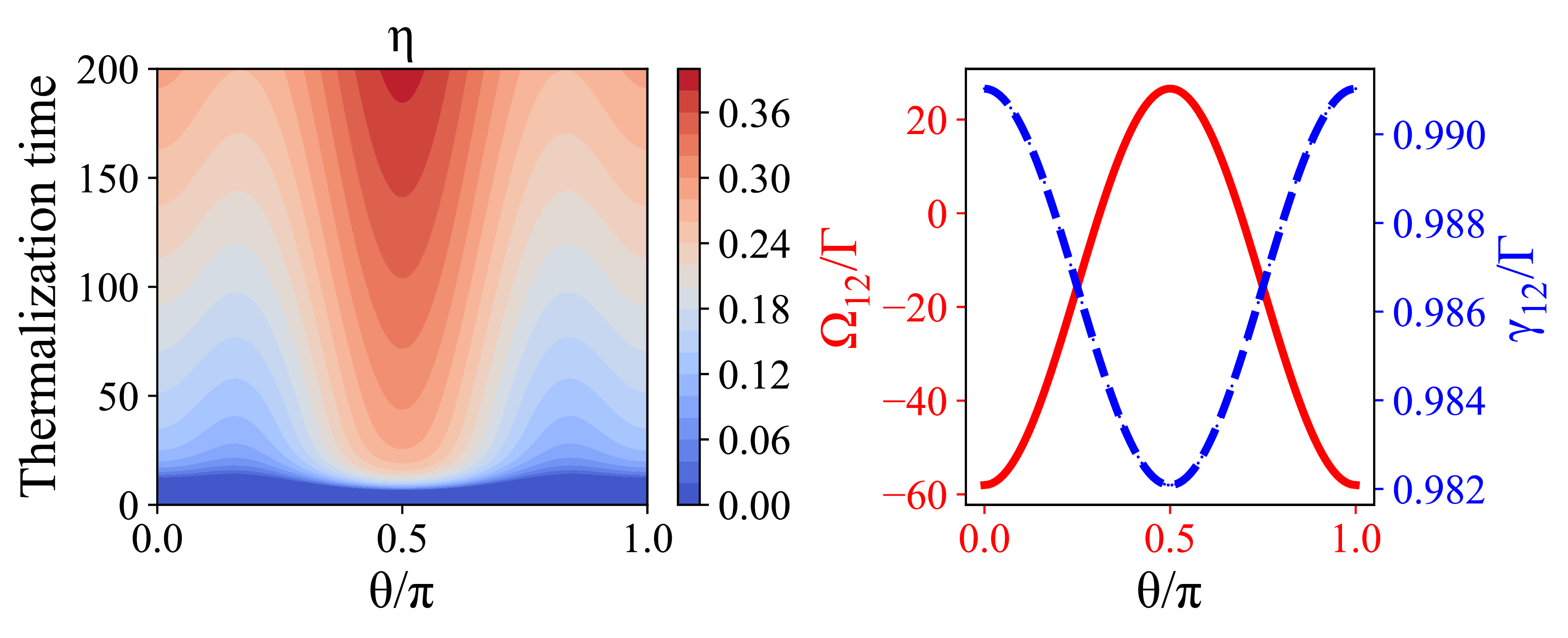}
 \caption{Trapped-atom engine operation for different dipole orientations ($\theta\in[0,\pi]$) and thermalization times. Heat absorbed from the hot reservoir, total work done by the engine, operating efficiency, and light-induced DDIs are respectively shown from (a) to (d) for $\xi=0.25$, and (e)to (h) for $\xi=0.3$. The other parameters are the same as in Fig. \ref{f2}.}\label{f88} 
\end{center}
\end{figure}

In Figs. \ref{f8}(a-d) and Figs. \ref{f88}(a-h), we explore the influence of changing the dipole orientation angle from $\theta=0$ to $\theta=\pi$ on the engine's thermodynamic behavior. This modification alternately affects the collective dissipations and frequency shifts, which in turn impact the engine's thermodynamic quantities. Notably, the frequency shift exhibits a sensitive variation, while the collective dissipation is only marginally affected in these figures. The system does not function as a heat engine in the parameter regions shaded in dark blue in Fig. \ref{f8}(b). However, efficient engine operation is achieved within the range of the dipole orientation $\theta\in[0.16\pi,0.84\pi]$. The high efficiencies observed in the limiting cases do not imply a high work output compared to perpendicularly oriented dipoles; rather, they indicate low heat absorption from the reservoir and work output. Interestingly, an optimum work output with a considerable efficiency is attained at $\theta=\pi/2$ as shown in Fig. \ref{f8}(c) (also see Figs. \ref{f88}(c) and \ref{f88}(g) below).
Furthermore, regardless of the chosen dipole orientation angle within the range of $0<\theta<\pi$, engine operation is also observed in Figs. \ref{f88}(a-d) for $\xi=0.25$ and in Figs. \ref{f88}(e-h) for $\xi=0.3$. The decreased engine efficiency in these parameter regimes can be attributed to the reduced interatomic distance, which strongly influences the collective frequency shift and dissipation, as explained in Fig. \ref{f2}. It is important to note that the engine performs poorly at $\xi=0.3$ compared to the cases when $\xi=0.2$ and $\xi=0.25$ in terms of efficiency, and increasing the thermalization time does not lead to an improvement in efficiency. The presence of collective dissipations remains the main cause for the low efficiency and high thermalization. Moreover, in these parameter regimes, the efficiency of the engine is lower, and the thermalization time required to initiate engine operation is approximately four times longer compared to that of a non-interacting working medium (see Figs. \ref{f3}(d) and \ref{f3}(h) in the main text). As the collective dissipations slowly vary and the collective frequency shift significantly decreases afterwards, there is no reasonable expectation for an improved efficiency compared to the non-interacting efficiency limit \cite{ar8}.
\begin{figure}
\begin{center}
\includegraphics[width=.45\textwidth]{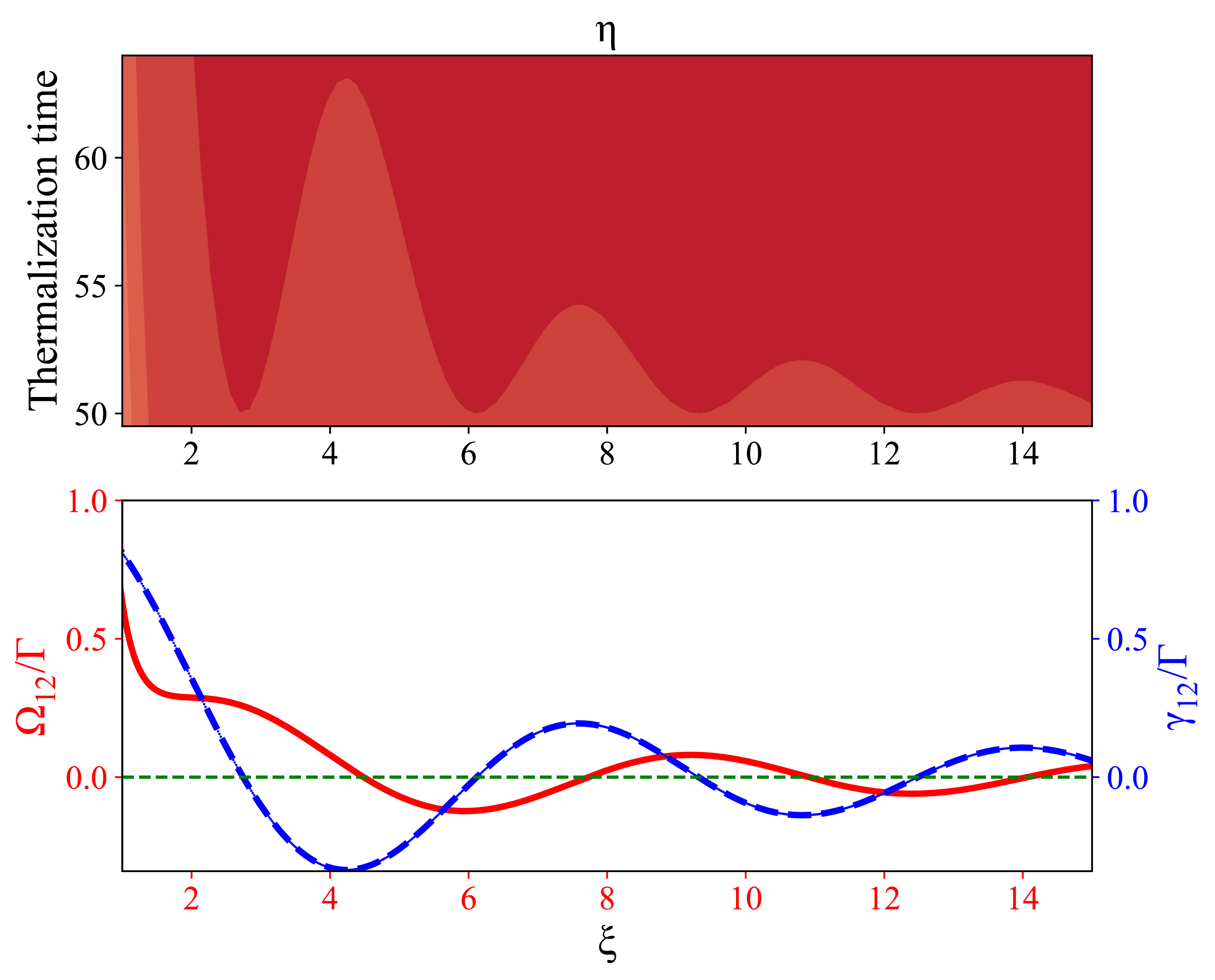}\\
 \caption{Enlarged part of maximum oscillating efficiency shown in Fig. \ref{f2}(e) and the corresponding light-induced DDIs. The other parameters are the same as in Fig. \ref{f2}.}\label{f9}
\end{center}
\end{figure}
\begin{figure}
    \centering
    \hspace{-2.5cm}(a)\hspace{3.5cm}(b)\\
    \includegraphics[width=.4\textwidth]{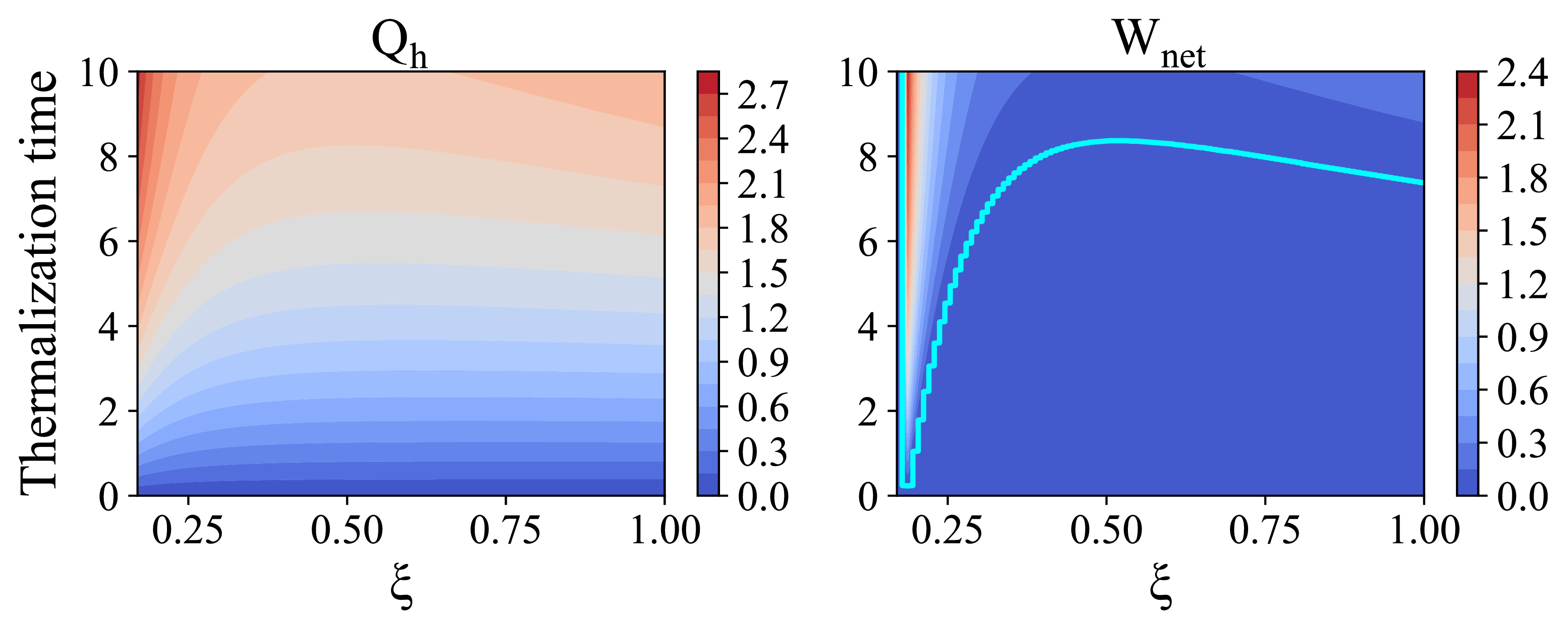}\\
    \hspace{-2.5cm}(c)\hspace{3.5cm}(d)\\
    \includegraphics[width=.4\textwidth]{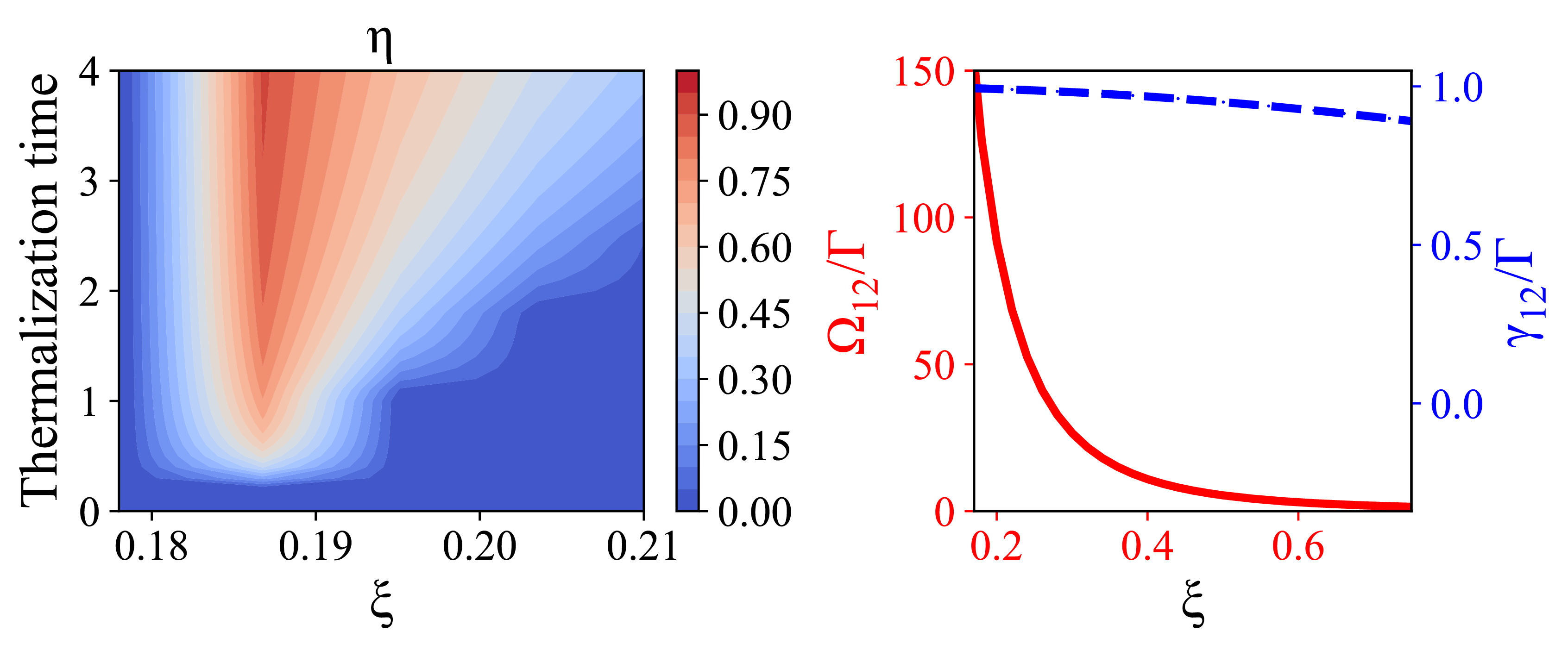}\\
    \hspace{-2.5cm}(e)\hspace{3.5cm}(f)\\
    \includegraphics[width=.4\textwidth]{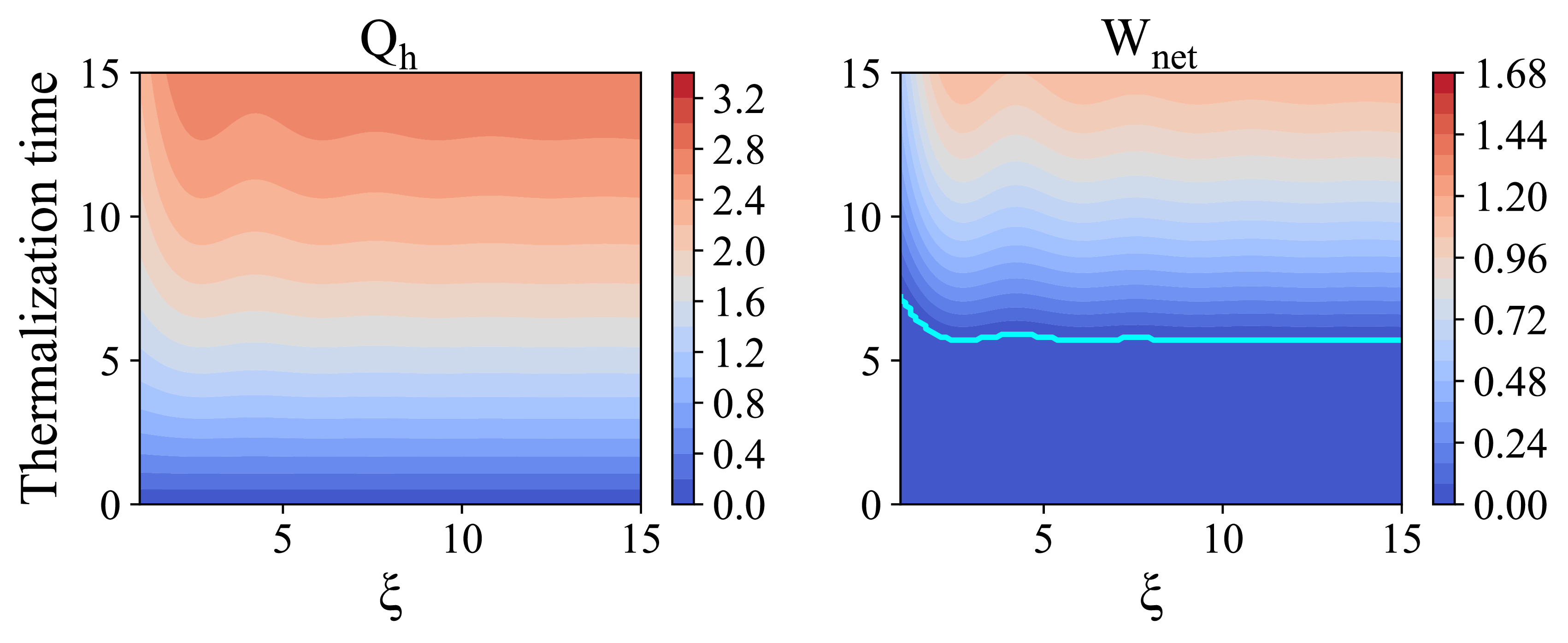}\\
    \hspace{-2.5cm}(g)\hspace{3.5cm}(h)\\
    \includegraphics[width=.4\textwidth]{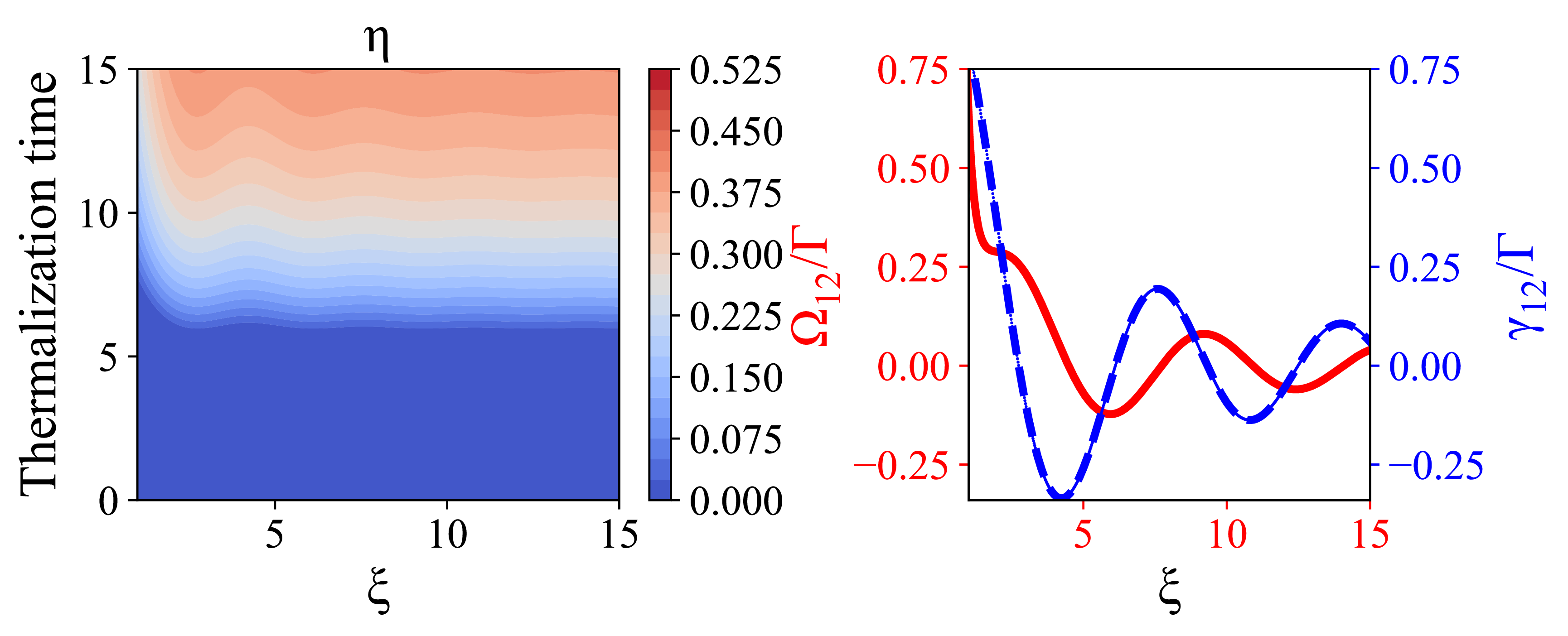}
    \caption{Trapped-atom engine operation for different interatomic distances ($\xi\in[0.175,15]$) and thermalization times. Panels (a) to (d) display the heat absorbed from the hot reservoir, total work done by the engine, operating efficiency, and light-induced DDIs, respectively, for $\xi\in[0.175,1]$. Panels (e) to (h) show the corresponding results for $\xi\in[1,15]$. The rest parameters are the same as in Fig. \ref{f2}.}\label{f10}
\end{figure}

As illustrated in Fig. \ref{f2}, the heat absorbed by the working medium, net work done, and efficiency exhibit oscillatory-like behavior as $\xi$ changes. These oscillations become prominent over extended thermalization times and are impacted by other system parameters including frequency shift. We compare the efficiency, frequency shift, and collective dissipation, as shown in Fig. \ref{f9}. Our analysis suggests that oscillations in thermodynamic quantities are solely induced by collective dissipations since peaks in the oscillating quantities correspond to vanishing collective dissipations, while valleys are achieved at the minimum or maximum values of collective dissipations. Therefore, understanding and controlling this dissipation allows to optimize the performance of the engine throughout the thermalization process.

In Figs. \ref{f10}(a-h), we present short-time behavior of thermodynamic quantities depicted in Fig. \ref{f2}. Collective frequency shift enables the engine to operate with nearly perfect efficiency by overcoming the effect of collective dissipations as shown in Figs. \ref{f10}(a-c). More importantly, for relative interatomic spacing within the range $\xi\in[0.18,0.195]$, thermalization time decreases by more than 60-fold compared to weakly interacting or non-interacting working medium (see Figs. \ref{f10}(b) and \ref{f10}(f)). However, for smaller $\xi$ values, the collective frequency shift abruptly distorts the atomic energy levels, rendering the system ineffective as an engine. This behavior has also been demonstrated under quasi-static operation of trapped-ion engine \cite{ar8}, where the driving field $g$ is not considered. In such cases, engine operation fails when $\Omega_{12}>B_h$. With an increase in interatomic spacing, we observe a rapid decrease in $\Omega_{12}$, accompanied by a gradual reduction in collective dissipation. Consequently, the thermalization time increases significantly compared to the non-interacting regime. We also show in the main text that the system undergoes a rapid thermalization process when the interatomic spacing is large, as demonstrated in Figs. \ref{f100}(a) and \ref{f100}(b). Therefore, in order to harness the quantum advantage of collective effects in our system, it is essential to fulfill both the requirements of finite-time operation and strong DDIs.

\begin{figure}
\begin{center}
\hspace{-2.75cm}{(a)\hspace{3.5cm}(b)}\\
\includegraphics[width=.45\textwidth]{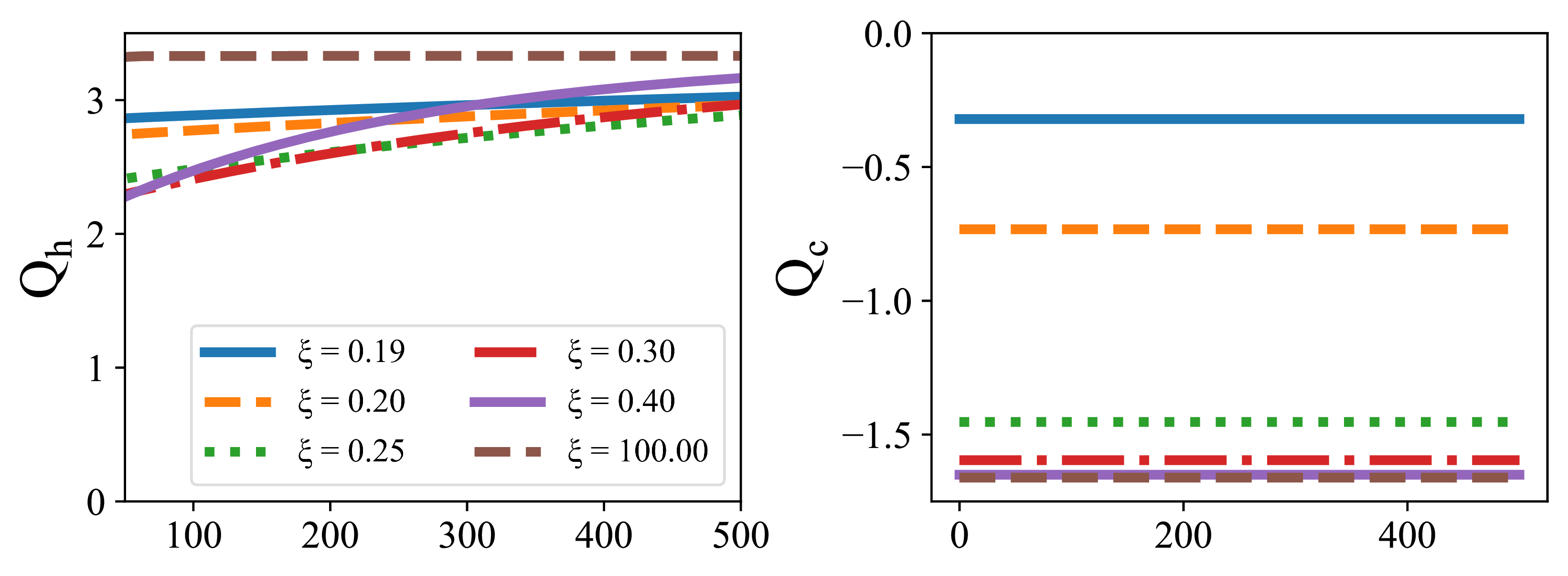}\\
\hspace{-2.75cm}{(c)\hspace{3.5cm}(d)}\\
\includegraphics[width=.45\textwidth]{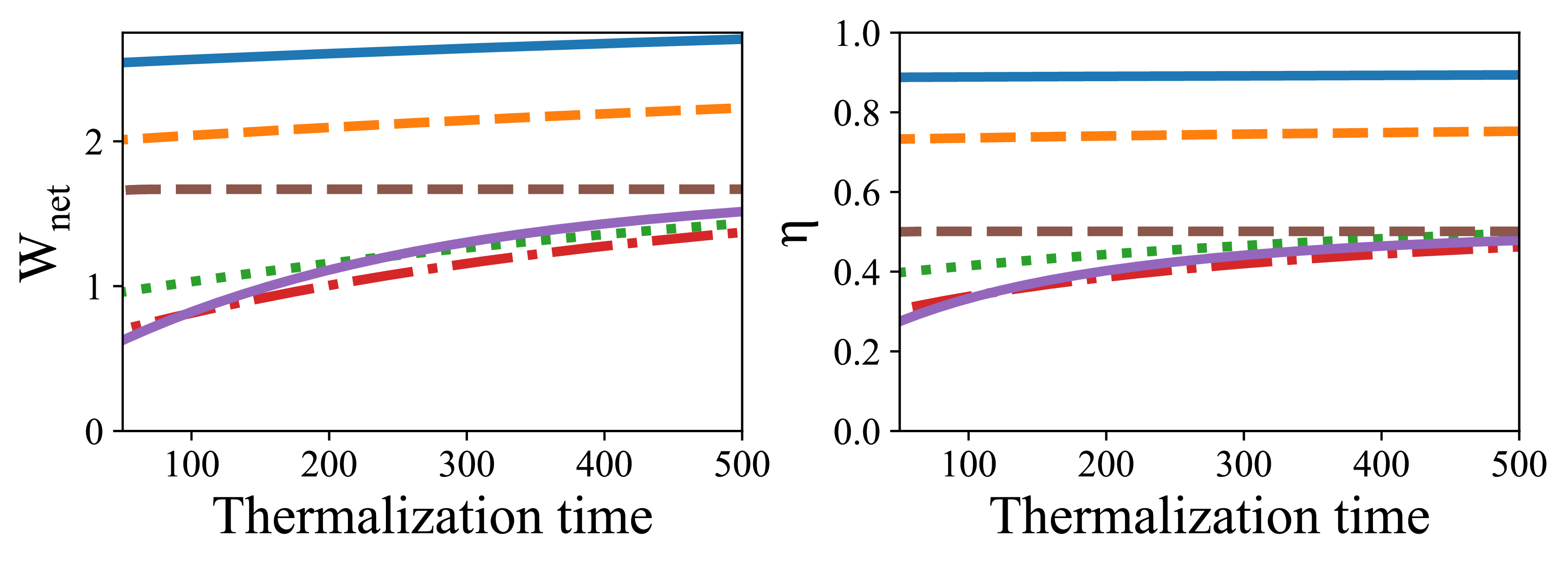}\\
\hspace{-2.75cm}{(e)\hspace{3.5cm}(f)}\\
\includegraphics[width=.45\textwidth]{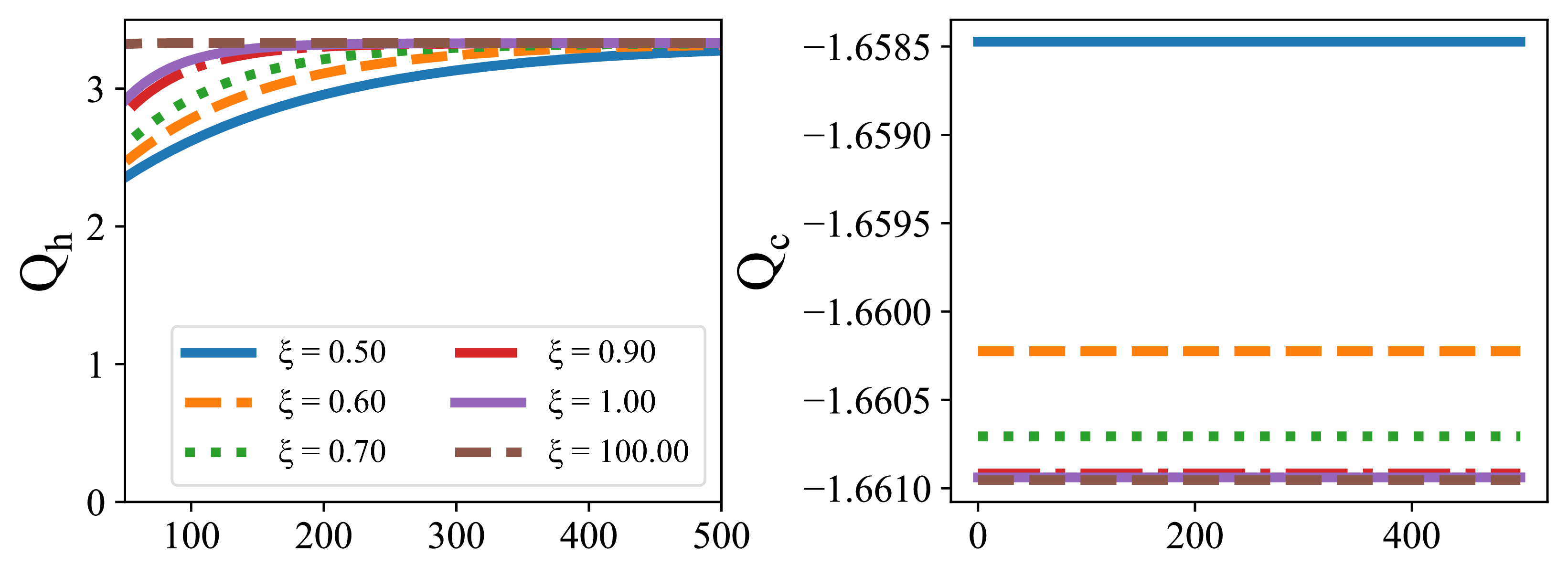}\\
\hspace{-2.75cm}{(g)\hspace{3.5cm}(h)}\\
\includegraphics[width=.45\textwidth]{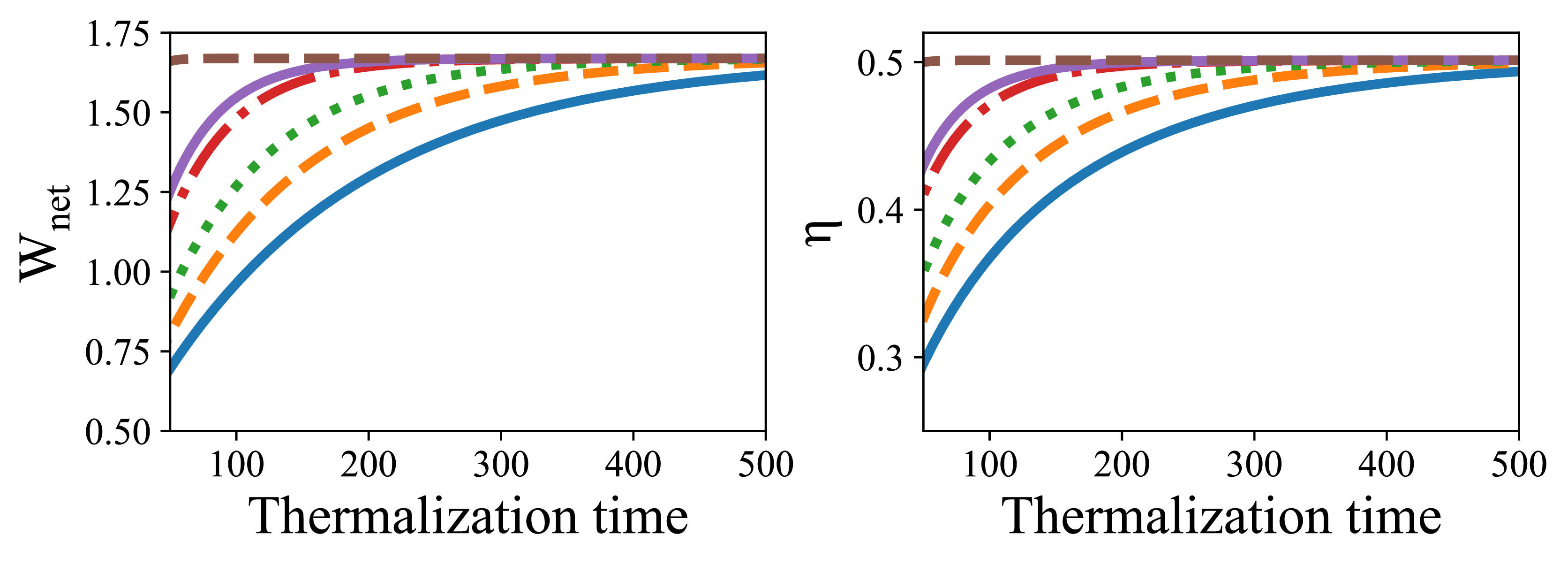}
 \caption{Trapped-atom engine operation for various interatomic distances and thermalization times. Panels (a) to (d) depict the heat absorbed from the hot reservoir, heat released to the cold reservoir, total work done by the engine, and operating efficiency, respectively, for different relative interatomic distances $\xi=0.19$ (solid-blue curve), $\xi=0.2$ (dashed-yellow curve), $\xi=0.25$ (dotted-green curve), $\xi=0.3$ (dash-dotted red curve), $\xi=0.4$ (solid-indigo curve), and $\xi=100$ (dashed-maroon curve). Panels (e) to (h) display the corresponding results for $\xi=0.5,0.6,0.7,0.9, 1$, and $100$. The rest parameters are the same as in Fig. \ref{f2}.}\label{f11}
\end{center}
\end{figure}

The performance of the engine in the long thermalization time is shown in Fig. \ref{f11}, which extends the results presented in Fig. \ref{f3} in the main text. Notably, as the thermalization time increases sufficiently, the plotted thermodynamic quantities tend to converge towards a single quasi-static value. This convergence is contingent upon the collective dissipation governed by the interatomic distances. Specifically, when the interatomic distances are smaller, a longer thermalization time is required for the thermodynamic quantities to reach their quasi-static counterparts.  
\section{\label{B}Finite thermalization and unitary dynamics}
\begin{figure}
    \centering
    \hspace{-2.25cm}(a)\hspace{3.5cm}(b)\\
    \includegraphics[width=.45\textwidth]{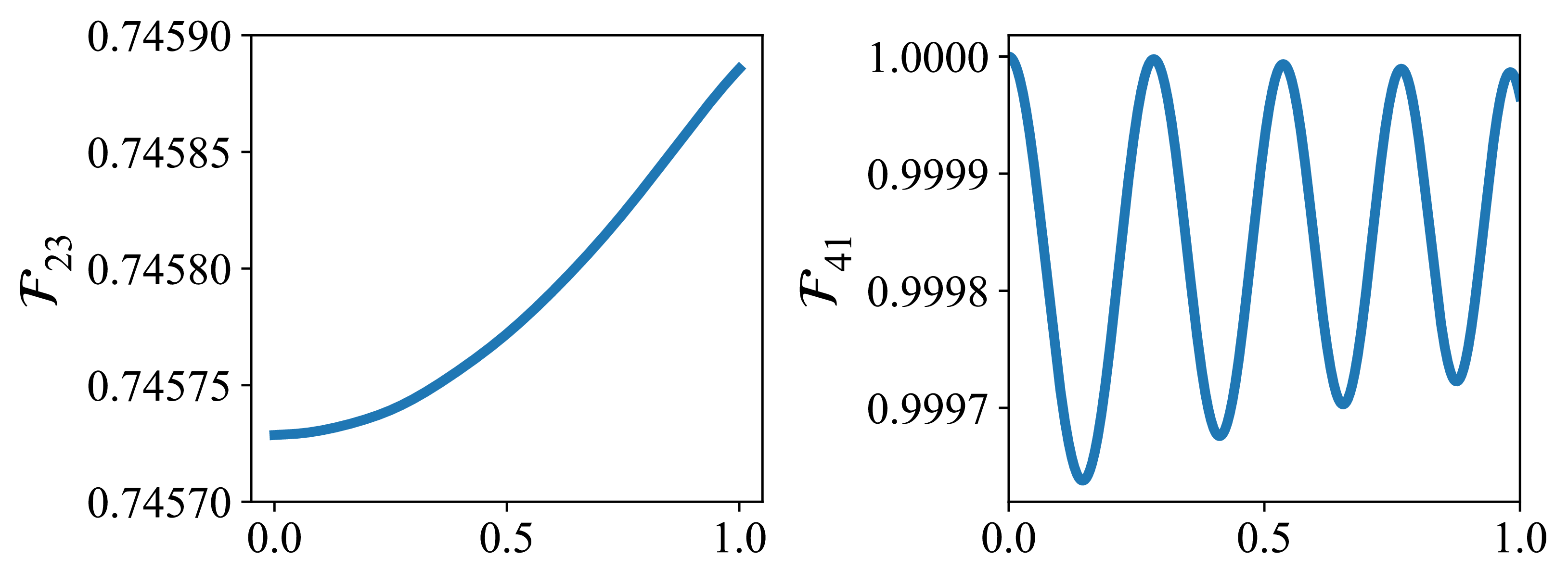}\\
    \hspace{-2.25cm}(c)\hspace{3.5cm}(d)\\
    \includegraphics[width=.45\textwidth]{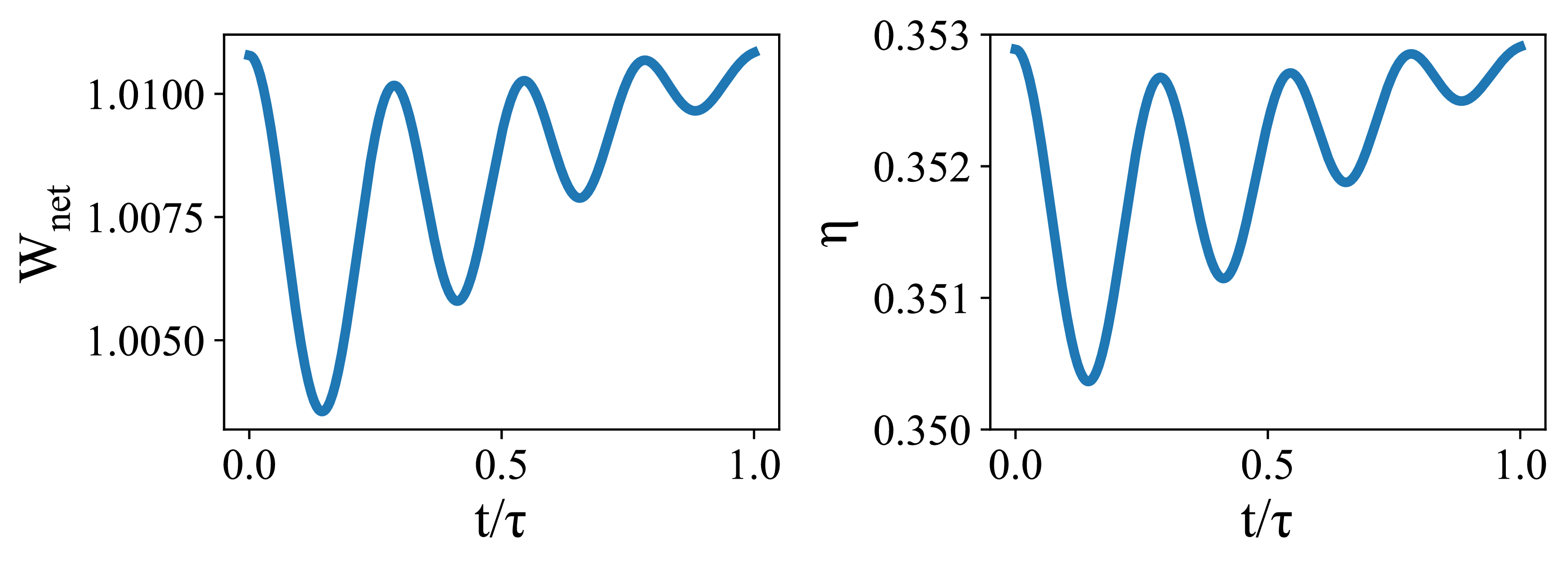}\\
    \hspace{-2.25cm}(e)\hspace{3.5cm}(f)\\
    \includegraphics[width=.45\textwidth]{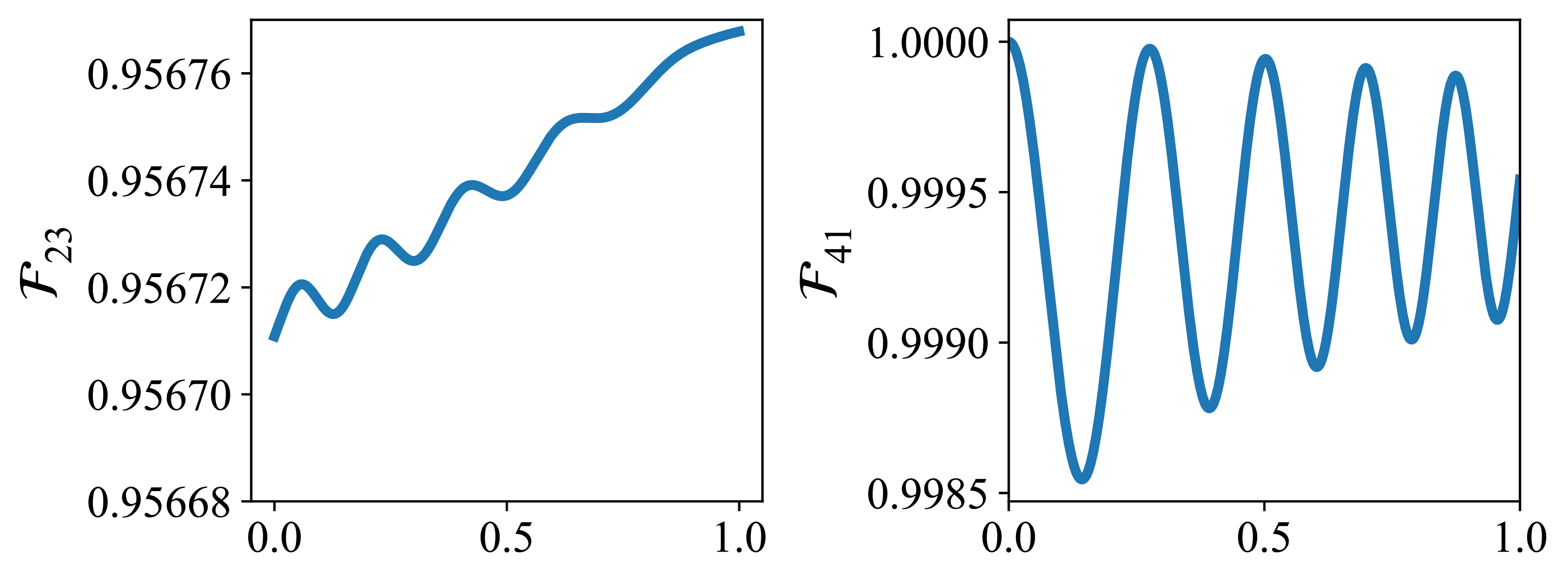}\\
    \hspace{-2.25cm}(g)\hspace{3.5cm}(h)\\
    \includegraphics[width=.45\textwidth]{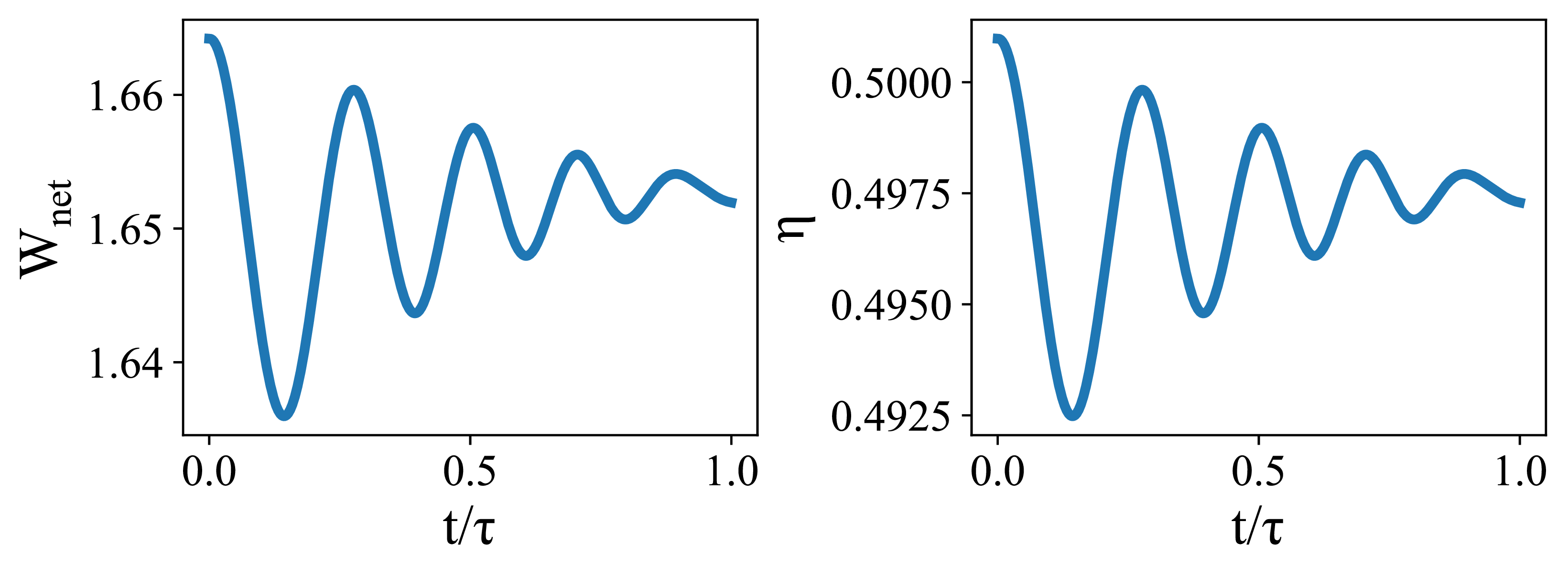}
    \caption{Trapped-atom engine operation for various interatomic distances and unitary driving times. Fidelity of second and forth strokes, total work done by the engine, and operating efficiency are, respectively, depicted from (a) to (d) for $\xi=0.19$. Panels (e) to (h) display the corresponding results for $\xi=100$. The unitary evolution lasts for $\tau=2/\omega$, and the rest parameters are the same as in Fig. \ref{f2}.}
    \label{f12}
\end{figure}
In Figs. \ref{f12}(a-h), we further clarify finite-time operation of the engine shown in Figs. \ref{f5} and \ref{f6} in the main text. We specifically choose the interatomic distances $\xi=0.19$ and $\xi=100$ as examples. The system exhibits low fidelity indicating that it is far from an equilibrium condition while still functioning as a heat engine. This departure from equilibrium can be seen in Fig. \ref{f10} and is discussed in detail in Section \ref{IV} A and B. The fidelity of the second stroke increases with the duration of the unitary driving, suggesting that short driving strongly excites the working medium and leads to the dissipation of energy along the driving trajectory. Furthermore, we show in Fig. \ref{f12}(b) that the excitation of the system during the fourth stroke is reduced for small interatomic spacing. However, coherence effects arising from finite thermalization and unitary driving strongly influence the engine performance during the second stroke depicted in Figs. \ref{f12}(c) and \ref{f12}(d). This performance is comparatively lower than that achieved under complete thermalization and adiabatic conditions discussed in Section \ref{IV} A and B. Nevertheless, both the work output and efficiency increase with the duration of the unitary driving. Similar observations apply to the case of nearly non-interacting atoms shown in Figs. \ref{f12}(e-h) for $\xi=100$. In this scenario, the system is closer to equilibrium since non-interacting systems quickly thermalize as observed in Figs. \ref{f100}(a) and \ref{f100}(b). As a result, the overall performance of the engine is good for non-interacting working medium compared to that of strongly interacting ones (see Figs. \ref{f12}(d) and \ref{f12}(h)). For instance, the efficiency near the sudden limit is approximately $35\%$ for $\xi=0.19$, whereas it reaches around $50\%$ for $\xi=100$.

\bibliography{apssamp}

\end{document}